\documentclass[%
 reprint,
 amsmath,amssymb,
 aps,
]{revtex4-2}

\usepackage{graphicx}% Include figure files
\usepackage{dcolumn}% Align table columns on decimal point
\usepackage{bm}% bold math
\usepackage{amsmath}
\usepackage{amssymb}
\usepackage[utf8]{inputenc}
\usepackage[T1]{fontenc}
\usepackage{mathptmx}
\usepackage{xcolor}
\usepackage{enumitem}   
\usepackage{subfig}
\usepackage[format=plain,labelfont=up,textfont=up]{caption}
\usepackage{dsfont}
\usepackage{xr-hyper}
\usepackage{hyperref}
\usepackage{url}
\usepackage{siunitx}
\usepackage{float}
\usepackage{soul,xcolor}

\usepackage{textgreek}

\begin{document}

\title{Exploring Functional Photonic Devices made from a Chiral Metal-Organic Framework Material by a Multiscale Computational Method}

\author{Benedikt Zerulla}
\affiliation{ Institute of Nanotechnology, Karlsruhe Institute of Technology (KIT), D-76344 Eggenstein-Leopoldshafen, Germany}
\author{Chun Li}
\affiliation{ Institute of Functional Interfaces, Karlsruhe Institute of Technology (KIT), D-76344 Eggenstein-Leopoldshafen, Germany}
\author{Dominik Beutel}
\affiliation{ Institute of Theoretical Solid State Physics, Karlsruhe Institute of Technology (KIT), D-76131 Karlsruhe, Germany }
\author{Simon Oßwald}
\affiliation{Institute of Organic Chemistry, Karlsruhe Institute of Technology (KIT), D-76131 Karlsruhe, Germany} 
\author{Christof Holzer}
\affiliation{ Institute of Theoretical Solid State Physics, Karlsruhe Institute of Technology (KIT), D-76131 Karlsruhe, Germany}
\author{Jochen Bürck}
\affiliation{Institute of Biological Interfaces (IBG-2), Karlsruhe Institute of Technology (KIT), D-76344 Eggenstein-Leopoldshafen, Germany}
\author{Stefan Bräse}
\affiliation{Institute of Organic Chemistry, Karlsruhe Institute of Technology (KIT), D-76131 Karlsruhe, Germany}
\affiliation{Institute of Biological and Chemical Systems–Functional Molecular Systems, Karlsruhe Institute of Technology (KIT), D-76344 Eggenstein-Leopoldshafen, Germany}
\author{Christof Wöll}
\affiliation{Institute of Functional Interfaces, Karlsruhe Institute of Technology (KIT), D-76344 Eggenstein-Leopoldshafen, Germany }
\author{Ivan Fernandez-Corbaton}
\affiliation{Institute of Nanotechnology, Karlsruhe Institute of Technology (KIT), D-76344 Eggenstein-Leopoldshafen, Germany }
\author{Lars Heinke}
\affiliation{Institute of Functional Interfaces, Karlsruhe Institute of Technology (KIT), D-76344 Eggenstein-Leopoldshafen, Germany}
\email{lars.heinke@kit.edu}
\author{Carsten Rockstuhl}
\affiliation{Institute of Nanotechnology, Karlsruhe Institute of Technology (KIT), D-76344 Eggenstein-Leopoldshafen, Germany}
\affiliation{Institute of Theoretical Solid State Physics, Karlsruhe Institute of Technology (KIT), D-76131 Karlsruhe, Germany }
\email{carsten.rockstuhl@kit.edu}
\author{Marjan Krsti\'c}
\affiliation{Institute of Theoretical Solid State Physics, Karlsruhe Institute of Technology (KIT), D-76131 Karlsruhe, Germany}
\email{marjan.krstic@kit.edu}


\begin{abstract}
Electronic circular dichroism is an important optical phenomenon offering insights into chiral molecular materials. On the other hand, metal-organic frameworks (MOFs) are a novel group of crystalline porous thin-film materials that provide tailor-made chemical and physical properties by carefully selecting their building units. Combining these two aspects of contemporary material research and integrating chiral molecules into MOFs promises devices with unprecedented functionality. However, considering the nearly unlimited degrees of freedom concerning the choice of materials and the geometrical details of the possibly structured films, we urgently need to complement advanced experimental methods with equally strong modeling techniques. Most notably, these modeling techniques must cope with the challenge that the material and devices thereof cover size scales from Ångströms to mm. In response to that need, we outline a computational workflow that seamlessly combines quantum chemical methods to capture the properties of individual molecules with optical simulations to capture the properties of functional devices made from these molecular materials. We concentrate on chiral properties and apply our work to UiO-67-BINOL MOFs, for which experimental results are available to benchmark the results of our simulations and explore the optical properties of cavities and metasurfaces made from that chiral material.   
\end{abstract}

\maketitle

\section{Introduction and Summary}
Electronic circular dichroism (ECD) spectroscopy is a stereoselective spectroscopy in the UV-Vis part of the spectrum, which is the chiroptical counterpart to the UV-Vis absorption. The method was pioneered in the second decade of the 19th century by François Arago \cite{Arago1811}, Jean-Baptiste Biot \cite{Biot1812, Biot1815, Biot1817, Biot1818}, and Augustin-Jean Fresnel \cite{AragoFresnel1819}. The ECD spectroscopy captures the difference in absorption between left- and right-handed circularly polarized light. While two enantiomers of a chiral molecule and material have the same UV-Vis absorption spectrum, their ECD spectra differ. Therefore, ECD spectroscopy is not a substitution but rather an important extension to the standard linear electronic spectroscopy that offers us more detailed insights into the structural and optical properties of molecular materials. Initial applications were in the studies of crystalline materials, fluids, and biologically vital chiral molecules \cite{Arago1811, Biot1812, Biot1815, Biot1817, Biot1818, AragoFresnel1819}. Today, we know many molecular materials that exhibit chiral properties, and a broad range of them have an application in medicine and industry. It is, therefore, extremely important to control the optical properties of novel materials and design devices that can be used to measure ECD spectroscopic properties. 

One example of a promising novel material that can benefit from a combination with chiral molecules are metal-organic frameworks (MOFs, a chiral metal-organic framework is also commonly abbreviated CMOF) \cite{James03,ZHUANG2016391} that can be prepared with a layer-by-layer liquid-phase epitaxy (LPE) method\cite{C7CS00315C} or grown via vapor-assisted conversion (VAC) \cite{doi:10.1021/jacs.7b08174}. This deterministic fabrication approach allows for precise growth, which results in a controlled thickness and orientation of the final thin film. MOFs are exponentially gaining popularity in the last two decades, and recently, the focus of their application shifted from catalytic \cite{Dandan2019, Bavykina2020} toward optical functionalities \cite{Wade2013,Santos2020,D0SC02436H}. In addition, the possibility to use virtually any organic molecule as a building block connecting metallic centers to design MOFs \cite{Haldar2019, Nefedov2021,https://doi.org/10.1002/adfm.201907625,https://doi.org/10.1002/adfm.202101335} has also drawn the focus to chiral MOFs. Furthermore, it is also possible to load achiral MOFs with chiral guest structures \cite{https://doi.org/10.1002/adfm.202204289}. The exciting opportunity to control the chirality of highly porous structured materials \cite{Gu2014, doi:10.1021/acs.chemrev.1c00740} intrigued researchers to explore applications in the fields of enantioselective chemistry and catalysis \cite{doi:10.1021/cs400959k, doi:10.1021/ja052431t, C4CS00094C}, chiral molecules separation \cite{doi:10.1021/jacs.9b07633, Seo2000}, optics \cite{https://doi.org/10.1002/adma.202002914, https://doi.org/10.1002/adom.202002096, https://doi.org/10.1002/adma.202101797, https://doi.org/10.1002/anie.201703917, C7CS00162B, doi:10.1021/cr200252n, Nefedov2021, Heinke2022, Mitric2022}, and sensors based on chiral MOFs \cite{doi:10.1021/ja302110d, doi:10.1021/jacs.0c13005, C5DT03283K, doi:10.1021/acsami.0c09578, Heinke2021}. 

However, the ongoing experimental efforts necessarily need to be complemented by accurate theoretical modeling\cite{Mitric2016}. The unprecedented opportunities to structure many different materials into a multitude of possible shapes to control the optical response make it indispensable to have a predictive modeling tool to study emerging properties before any fabrication. However, these developments turned out to be challenging since a complete understanding of the origin and control of underlying chiral properties, especially in devices, requires an interdisciplinary approach spanning across (quantum) chemistry, physics (optics), materials science, and engineering. Furthermore, the different length scales intrinsic to such devices, starting from the details of the individual molecule up to the entire device, prompt a holistic approach unavailable until now.

Here, we present a multiscale modeling workflow adapted to study the optical properties of functional devices made from chiral MOFs. The work significantly widens the scope of a previously developed approach \cite{SURMOFCavity} to the realm of chiral materials. We open a route to unveil basic physics governing light-chiral matter interactions with circularly polarized light, the possibility of tailoring targeted responses, and the knowledge-driven design of chiral devices (sometimes also called digital twins). In this context, we discuss exemplarily an optical cavity and a metasurface below. Our approach allows one to identify and assign the modes of cavities containing molecular materials \cite{Groenhof2019,Climent2020}.

We demonstrate our workflow using the example of Zirconium-based UiO-67 type MOF with the chiral [1,1'-Binaphthalene]-2,2'-diol (BINOL) organic linker built in the framework by a VAC method and study the electronic circular dichroism (ECD) spectroscopic properties of UiO-67-BINOL thin films. Molecular quantum chemistry simulations of MOF unit cells constitute the basis for obtaining the accurate optical response of the material to circularly polarized light. The connecting point between the spatial scales of the atomistic studies and the device simulations are the transition matrices (T-matrices)\cite{Fernandez-Corbaton:2020}. The T-matrices are constructed from dynamic polarizabilities calculated within the state-of-the-art time-dependent density functional theory (TD-DFT) framework and considered by full-wave Maxwell’s solvers. The efficient T-matrix approach for periodic systems allows us to effortlessly vary, in a later stage, geometrical details of functional devices that exploit these molecular materials in their design and to study emerging structure-property relations.

Our workflow consists of the following steps: 
\begin{enumerate}[label=(\roman*)]
\item Optimization of the unit cell of the crystalline MOF.
\item Defining a finite-size model capturing important optical interactions.
\item Calculating TD-DFT spectra to determine the spectral range of interest and understanding the origin of spectral features.
\item Calculating the electric-electric, electric-magnetic, and
magnetic-magnetic dynamic polarizabilities in dipole approximation for discrete frequencies in a given frequency range.
\item Constructing of T-matrices out of these polarizabilities.
\item In some cases\footnote{For specifically structured devices, such as the cylinders considered further below, we perform a homogenization of the molecular material based on the T-matrices to extract effective material parameters. This allows us to consider arbitrarily shaped objects made from a homogenous material.}: Homogenizing the MOF material based on the T-matrices.
\item Performing full-wave optical simulations of materials and devices based on scattering codes or finite-element methods.
\item Analyzing the results concerning the emerging optical functionality. 
\end{enumerate}

This manuscript is divided into the following sections to present all these details. Section 2 contains the results of the first five steps of the above-described workflow. These steps concern the quantum chemical details of the problem. Section 3 contains experimental and numerical simulation results of the MOF thin film material on a glass substrate. Section 4 depicts the power of our multiscale approach presenting the simulations of two photonic devices containing our novel chiral material: (a) a Fabry-Pérot cavity made from thin aluminum films where in between the UiO-67-BINOL MOF material is placed, and (b) a metasurface made from periodically arranged cylinders made by the same MOF material. For the optical cavity, we observe a splitting of the cavity modes for specific values of the MOF thin film thickness due to the interaction between the modes and the MOF material. For the metasurface, lattice effects between the chiral MOF cylinders enhance and even reverse the circular dichroism signal from the material, an example of a rather drastic modification of the intrinsic material properties upon structuring. Finally, we give a conclusion and a future perspective of our work.

\section{Quantum Chemistry Simulations of UiO-BINOL MOF}

This section outlines the most critical aspects of the quantum-chemical simulations. The purpose is to describe the bianisotropic properties, i.e., the properties that capture the details of the light-matter interaction of the MOF unit cell in a manner that the molecular material can be considered afterward in a more classical optical simulation. Therefore, we will initially optimize the geometry and compute the dipolar polarizability of the unit cell afterward.

\subsection{Structural properties of the porous crystalline UiO-67-\textit{R}-BINOL MOF}
The quantum mechanical and optical simulations within density functional theory (DFT) and its time-dependent variant (TD-DFT) method follow the approach for the workflow as described in Ref.~\cite{SURMOFCavity}. However, the novelty of the current work is the focus on a chiral MOF.
The workflow consists of three steps. 

\begin{figure*}
\centering
\subfloat{
\includegraphics[width=1.0\textwidth]{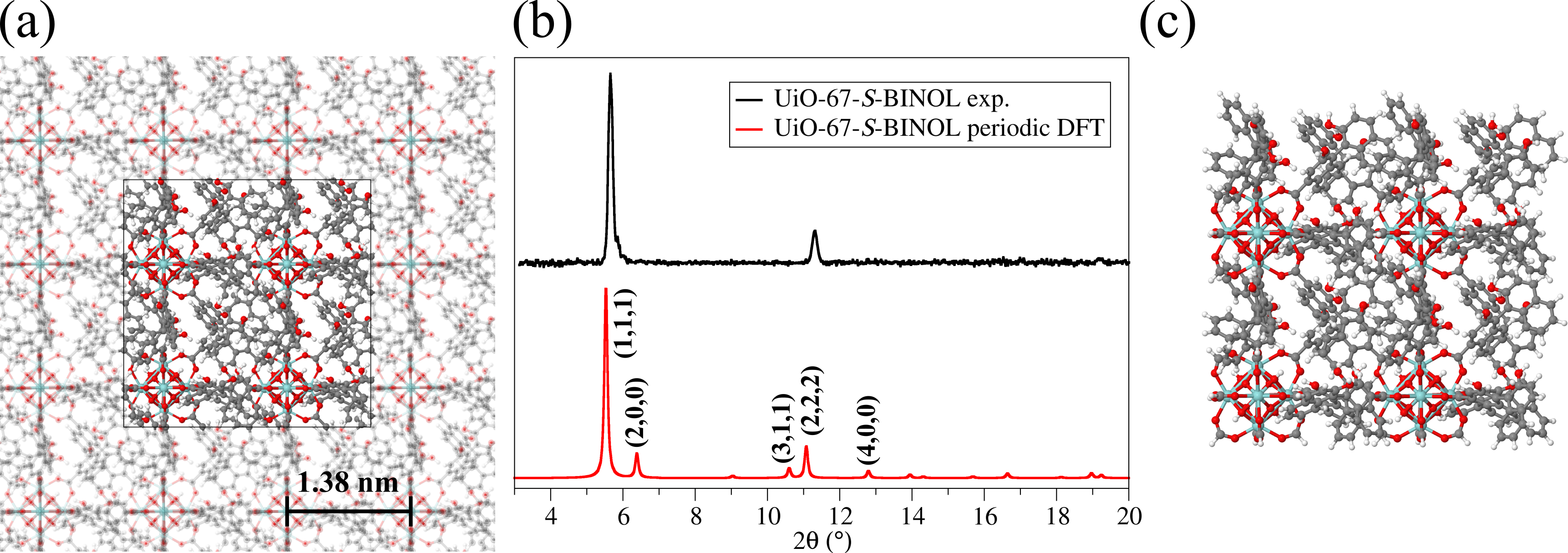}}\\
\subfloat{
    \includegraphics[width=0.36\textwidth]{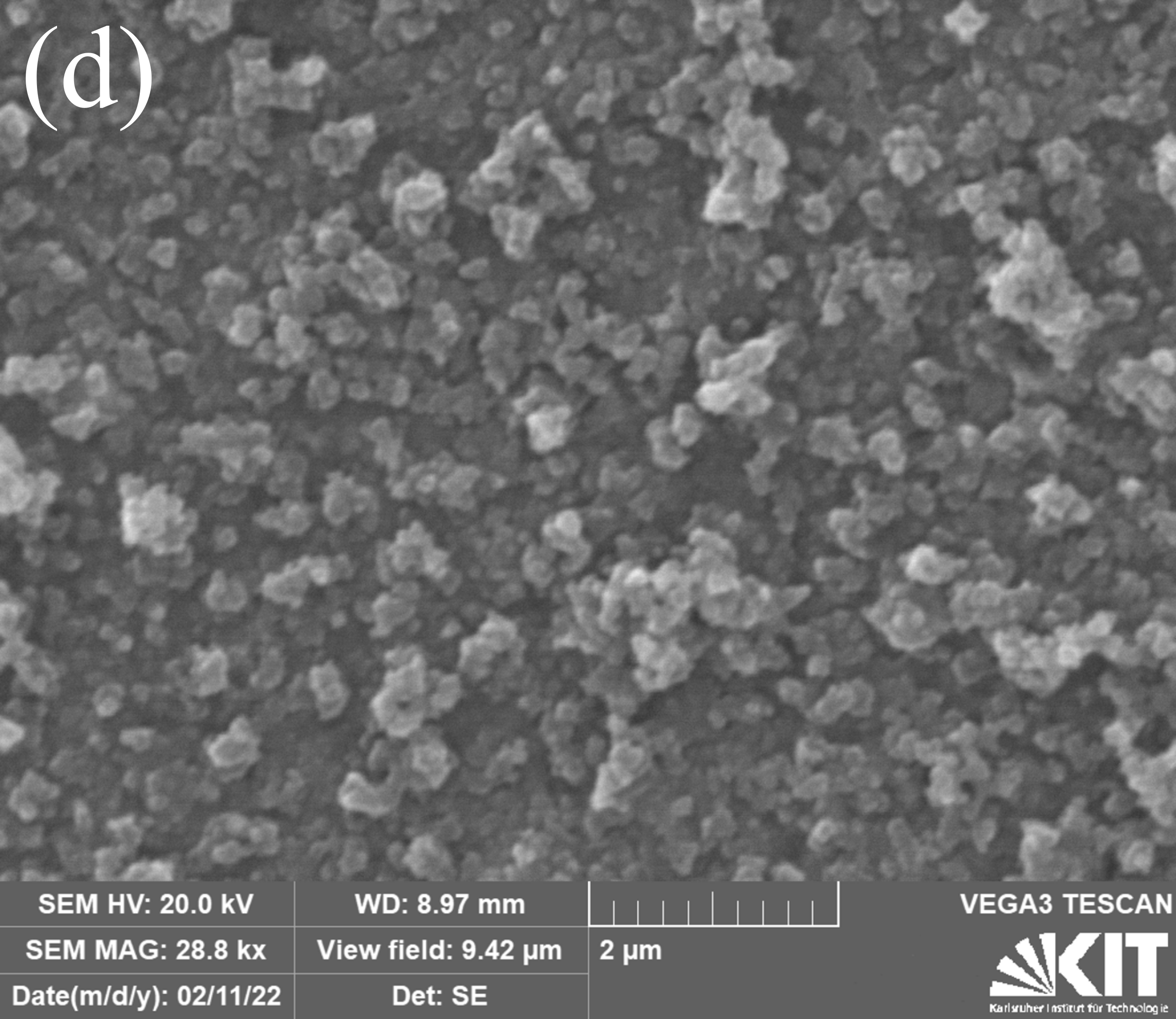}}
    \hspace{0.3 cm}
    \subfloat{
	\includegraphics[width=0.36\textwidth]{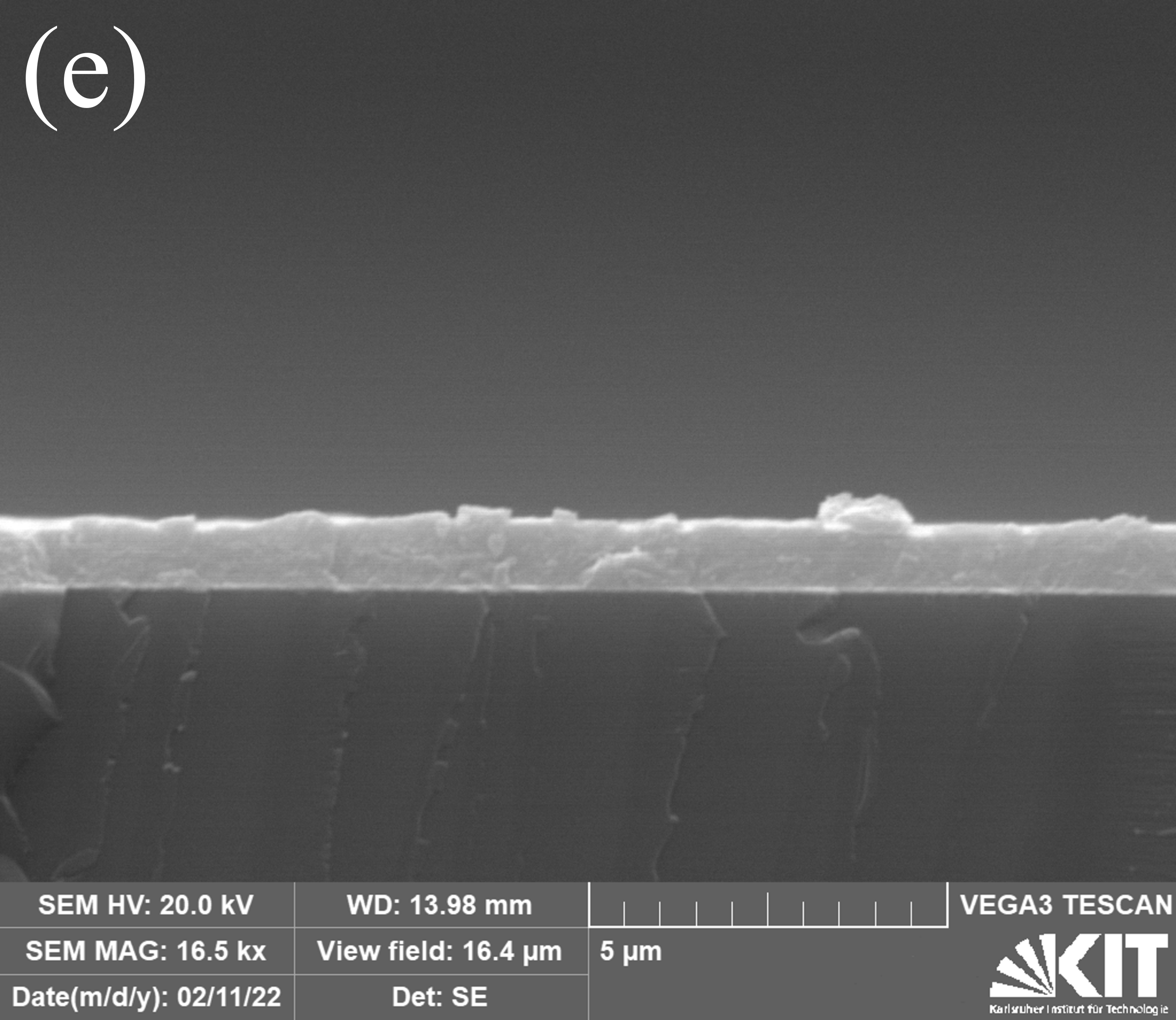}}
\caption{\textbf{(a)} The optimized structure of the unit cell of the periodic crystalline UiO-\textit{R}-BINOL MOF. \textbf{(b)} Comparison of simulated XRD pattern for the periodic DFT optimized structure of the UiO-\textit{R}-BINOL unit cell to the experimental XRD measurement of the UiO-\textit{S}-BINOL thin film. \textbf{(c)} Finite-size cluster-like molecular model of the UiO-\textit{R}-BINOL MOF as considered in the further calculation of the polarizabilities. In \textbf{(a)} and \textbf{(c)}, white, grey, red, and turquoise spheres represent H, C, O, and Zr atoms, respectively. The top-view \textbf{(d)} and side-view \textbf{(e)} SEM images of the prepared UiO-67-BINOL MOF thin film on the substrate.}
\label{fig:StructuresXRD}
\end{figure*}

First, we build and optimize the periodic porous UiO-\textit{R}-BINOL material using the periodic DFT code CP2K. The structure and unit cell dimensions were optimized simultaneously. Energetically the most optimal structure (Figure~\ref{fig:StructuresXRD}\textbf{(a)}) is characterized by a cubic unit cell with vector $||\bm{a}||= 27.647106$\,\textup{\AA}. The optimized structure of the unit cell was used to simulate the X-ray diffraction (XRD) pattern in the range $2\theta = 3^{\circ}-20^{\circ}$ for the incident wavelength of the X-rays $\lambda = 1.54$\,\textup{\AA} corresponding to the experimental conditions against which we compare our predictions (Figure~\ref{fig:StructuresXRD}\textbf{(b)}). Details of the experimental material preparation procedures, setups, and measurements can be found in the Supporting Information. 
In Figure~\ref{fig:StructuresXRD}\textbf{(b)}, it can be seen that the diffractions on the (111) and (222) planes are reproduced at the matching angles and with the same intensity ratio. Additional low-intensity peaks in the theoretical spectrum represent in-plane diffractions inaccessible for the out-of-plane XRD experimental measurements. The comparison between the experimental XRD pattern and the simulated one for the optimized material confirms that the identified structure is indeed the one that is experimentally accessible. These insights provide us with confidence to proceed with the subsequent optical calculations.
In the second step, we select and build a finite-size molecular model of the optimized periodic material from the first step. That is necessary to preserve a correct and complete electronic structure of the building blocks, in this case, ligands, of such chiral MOF material from which chiral properties originate. Ideally, the molecular model is as small as possible for computational efficiency. On the other hand, it needs to be large enough to capture all interactions critical for the optical properties of the MOF material. In the pertinent case, we consider the optimized unit cell in its entity, and the structure is rearranged so that all BINOL linker molecules located at the boundary of the unit cell of the MOF are not cut in half due to the periodic boundaries. This is important to ensure that the simulated optical response stems exclusively from the chiral components of the MOF. Furthermore, upon transitioning from the periodic description of the MOF to the cluster-like approach, we saturated the broken bonds between the BINOL linkers and the Zr-metallic centers with additional hydrogen atoms. This entire procedure yields a molecular model (Figure~\ref{fig:StructuresXRD}\textbf{(c)}) with 1074 atoms in total. This system was used to calculate dynamic complex polarizabilities in step iv) of the workflow.

\subsection{Dynamic polarizabilities and optical properties of UiO-67-BINOL MOFs}

\begin{figure*}
\centering
\includegraphics[width=1.0\textwidth]{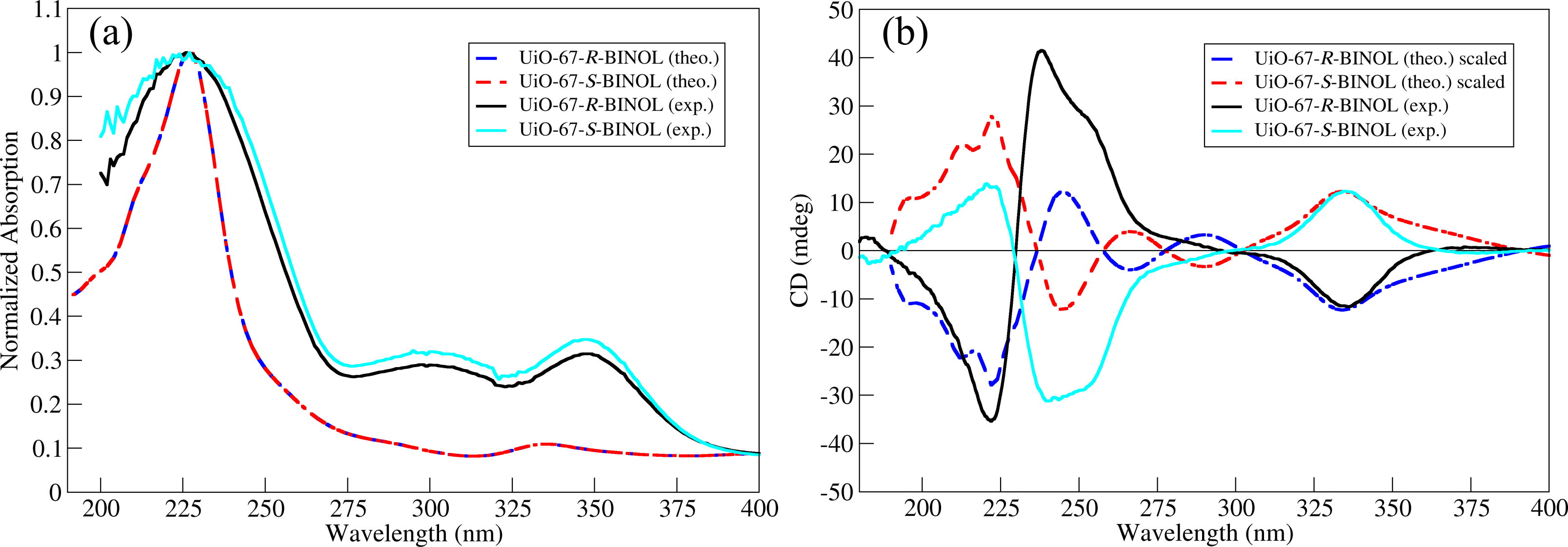}
\caption{\textbf{(a)} Comparison of TD-DFT simulated UV-Vis absorption spectra of the molecular model of UiO-BINOL MOF to experimental results of the MOF thin film for both enantiomers. \textbf{(b)} Comparison of simulated ECD spectra of the molecular model of UiO-BINOL MOF to experimental CD spectroscopic measurement of thin films for both enantiomers.}
\label{fig:TDDFTexpspectra}
\end{figure*}

The third step considers the molecular model from the previous step, and we proceeded by calculating the dynamic polarizabilities in dipole approximation using density functional theory (DFT) for the spectral range of interest (180-400\,nm). The spectral resolution was 2\,nm. The frequency-dependent dynamic polarizability tensors $\bm{\alpha}_{\nu \nu'} (\omega)$ are prerequisites to construct the T-matrices used in the optical simulations of the entire UiO-BINOL MOF thin film and devices thereof. The electric-electric ($\mathrm{ee}$), 
electric-magnetic ($\mathrm{em}$), and magnetic-magnetic ($\mathrm{mm}$) dipole polarizabilities can be obtained as in \cite{SURMOFCavity}
\begin{equation}
    \bm{\alpha}_{\nu \nu'} (\omega) = \langle \bm{x}^{\nu}, \bm{y}^{\nu'};  \bm{p}^{\nu'},\bm{q}^{\nu'} \rangle
\end{equation}
with the perturbations $\nu, \nu' \in \{ \mathrm{e}, \mathrm{m} \}$. $\bm{x}$ and $\bm{y}$ are the solutions of the 
coupled-perturbed Kohn--Sham equations. $\bm{p}^{\nu'}$ and $\bm{q}^{\nu'}$ represent the electric and magnetic dipole integrals 
transformed to the basis of the molecular Kohn--Sham orbitals \cite{SURMOFCavity}. Finally, we exploited that we can simulate the 
rotationally averaged UV-Vis absorption spectrum from the trace of the imaginary part of the calculated damped $\bm{\alpha}_{\mathrm{ee}}$ 
tensors at each discrete wavelength, and we use the trace of the real part of the damped $\bm{\alpha}_{\mathrm{em}}$ tensors to produce the 
ECD spectrum in the UV-Vis spectral range. The comparison of the spectra reconstructed from the damped dynamic polarizability tensors 
to those experimentally measured is shown in Figure~\ref{fig:TDDFTexpspectra}. Details of the measurements can be found in the Supporting Information. 

The theoretical UV-Vis absorption spectrum of both MOF enantiomers depicted in Figure~\ref{fig:TDDFTexpspectra}\textbf{(a)} predicts the experimental spectral features remarkably well. There are three broad peaks characterizing the absorption. Two low-intensity peaks at 350\,nm and 300\,nm are slightly blue-shifted in the simulations by \textasciitilde15\,nm while the most intensive band at 225\,nm is predicted on the exact position as experimentally observed. The experimental spectra are broadened almost twice compared to the chosen theoretical broadening of 0.3\,eV at full-width at half-maximum (FWHM). 

Comparing the simulated and experimental ECD spectra in Figure~\ref{fig:TDDFTexpspectra}\textbf{(b)}, the match is even more pronounced. The simulation does not only predict all resonance positions exactly but also the chosen broadening matches the experiments across the entire spectral region. The lowest lying peak at 334\,nm is used to scale the theoretical spectra for both enantiomers. The middle band predicted at 244\,nm in the theoretical spectrum is of the same magnitude as the lowest peak at 334\,nm while in the experimental measurements, it is of the same intensity as the peak at 222\,nm. Finally, the high energy peak at 222\,nm is the most intensive in the simulations and matches the shape of the measured spectrum perfectly in that spectral window. 

Once the polarizabilities are known, it is merely a change of basis and a rescaling to express the T-matrix\cite{Fernandez-Corbaton:2020}. All characteristic features of the spectra depicted in Figures~\ref{fig:TDDFTexpspectra}\textbf{(a)} and \textbf{(b)} are also present in the T-matrices in the considered spectral region. Some selected T-matrices are shown in Figure~S3 for illustration. With the T-matrices and these findings at hand, we can perform the transition from the quantum chemical properties and response of one unit cell to the optical properties and response of a MOF thin film.  

\section{Chiral UiO-BINOL MOF Thin Film on a Glass Substrate: Response of MOF with Rotationally Averaged Unit Cells}

In the next step, we perform multiscale simulations and consider the optical properties of a chiral UiO-BINOL MOF thin film on a glass substrate. In the Appendix, the T-matrix formalism and its usage for multiscale simulations are explained in more detail. In short, we can calculate, using that formalism, the reflection and transmission from a periodic planar arrangement of objects that are optically described by their T-matrices. Here, we consider the periodicity imposed by the MOF and the molecules considered above as individual objects. Stacking such layers is also possible so that we can predict the optical response from layers made from periodically arranged molecular unit cells. That approach allows for easily building entire MOF films with a macroscopic thickness. Besides layers containing periodically arranged molecules, interfaces and more traditional optical materials such as glasses or metals can be considered. In combination, that framework is perfectly suitable to explore the optical response of stratified media containing materials such as those considered here. 

Further below, we show the results of optical simulations adapted to realistic measurement scenarios and compare them to the corresponding experimental results. This section justifies the usage of the calculated T-matrices for different photonic devices presented in Section~\ref{Sec:Applications}.

\begin{figure*}
\centering
     \subfloat{
	\includegraphics[width=0.25\textwidth]{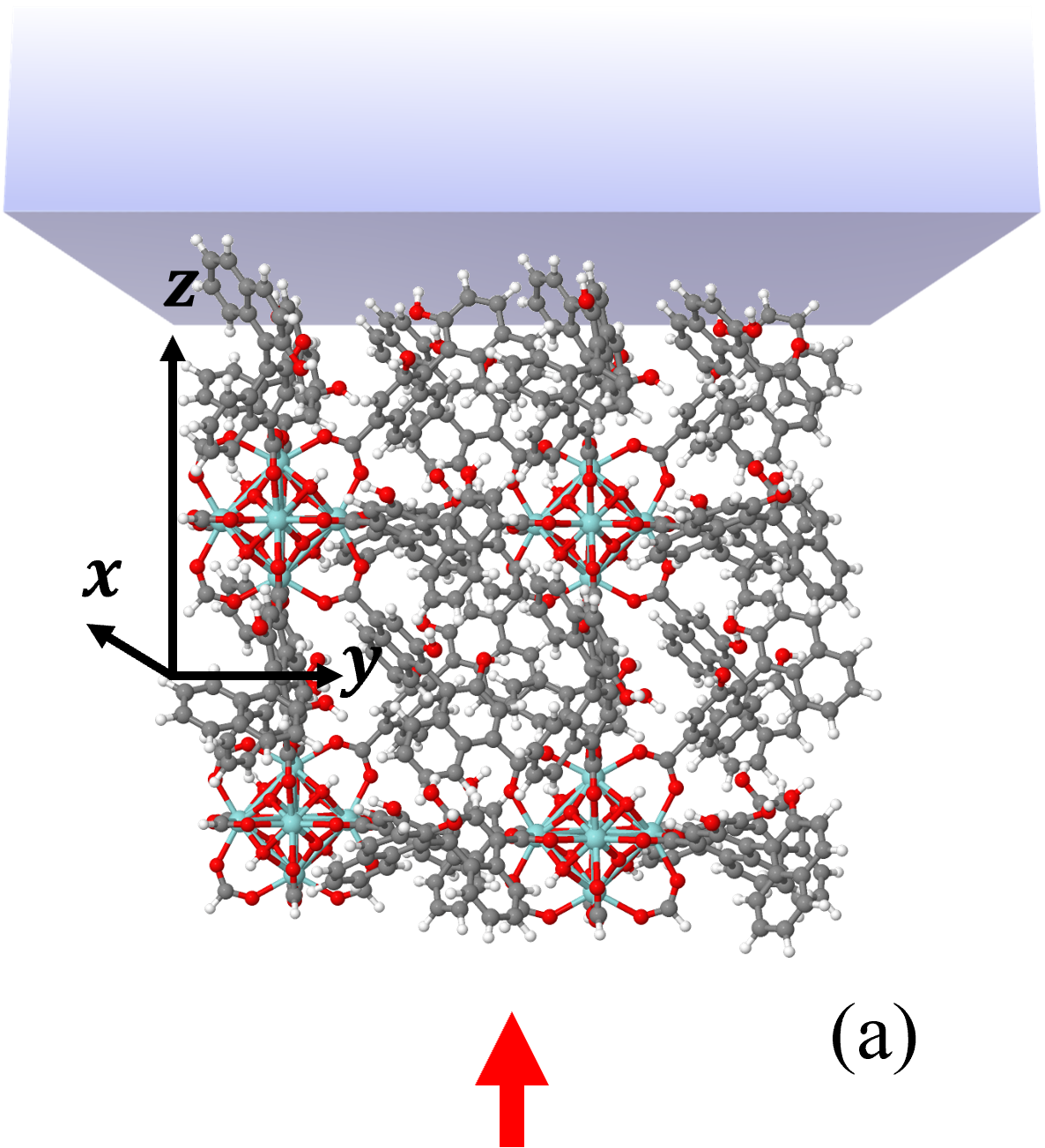}
 }
 \subfloat{
	 \includegraphics[width=0.225\textwidth]{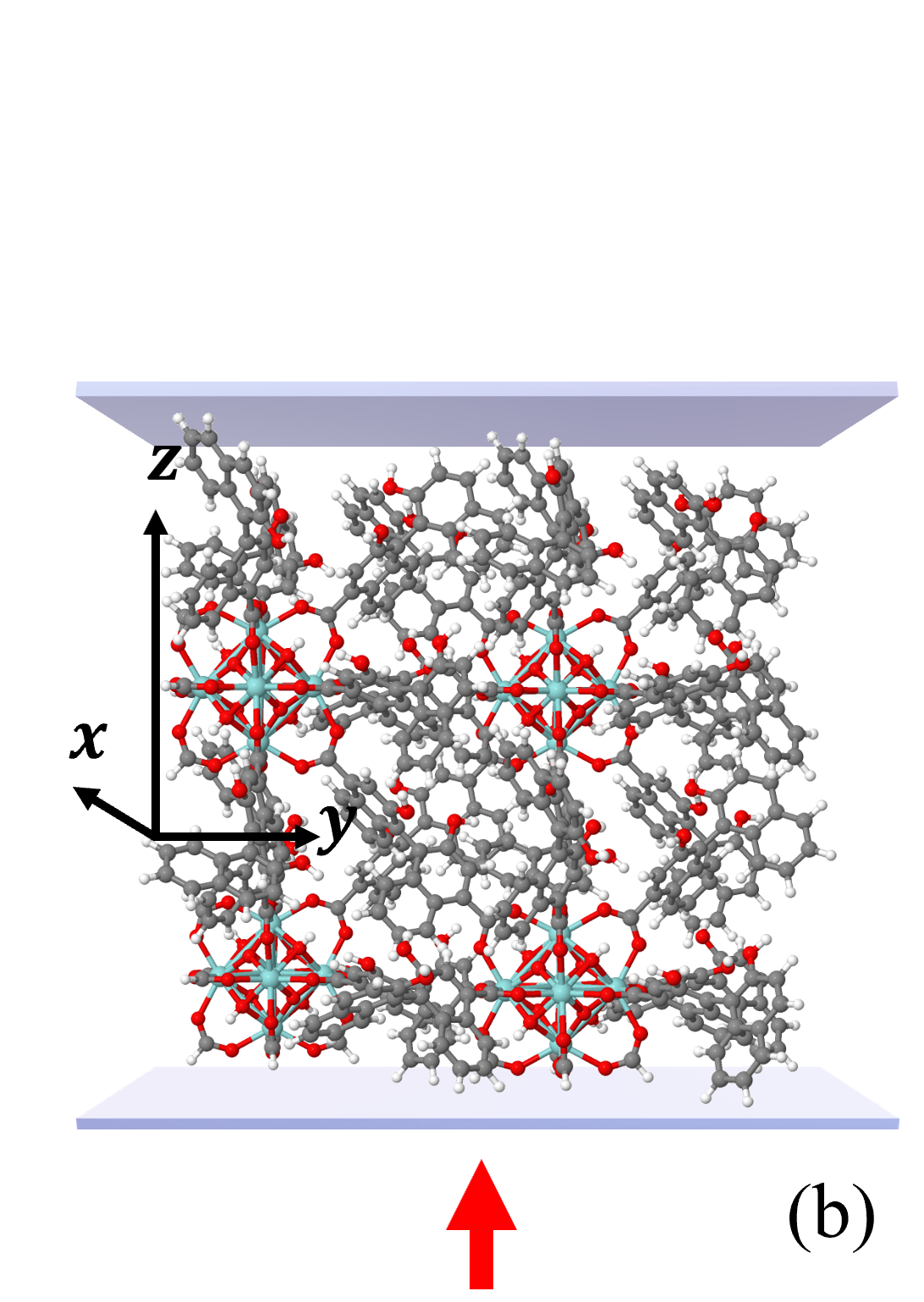}
  }\hspace{0.125cm}
  \subfloat{
	 \includegraphics[width=0.175\textwidth]{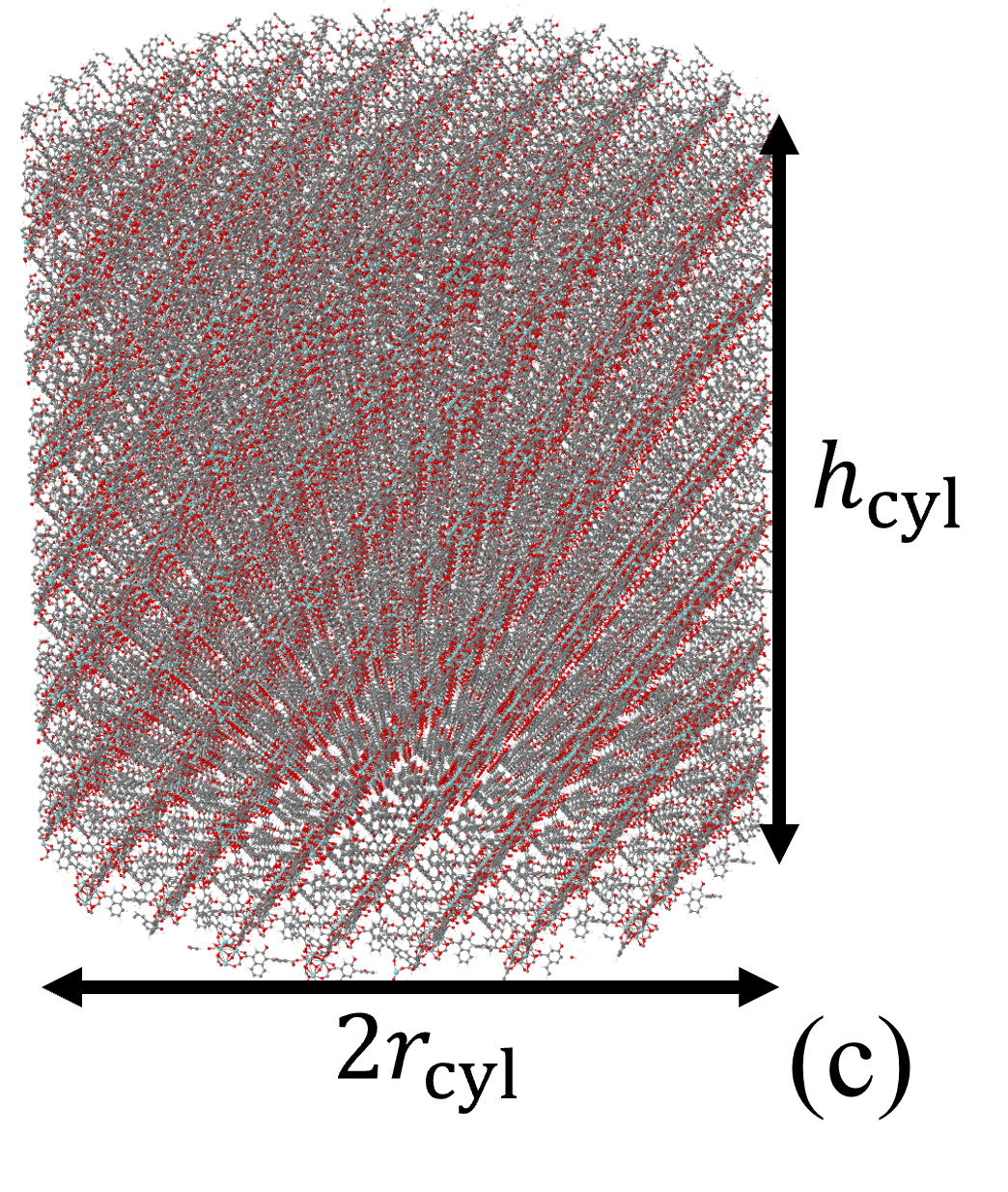}
  }
  \subfloat{
	 \includegraphics[width=0.3\textwidth]{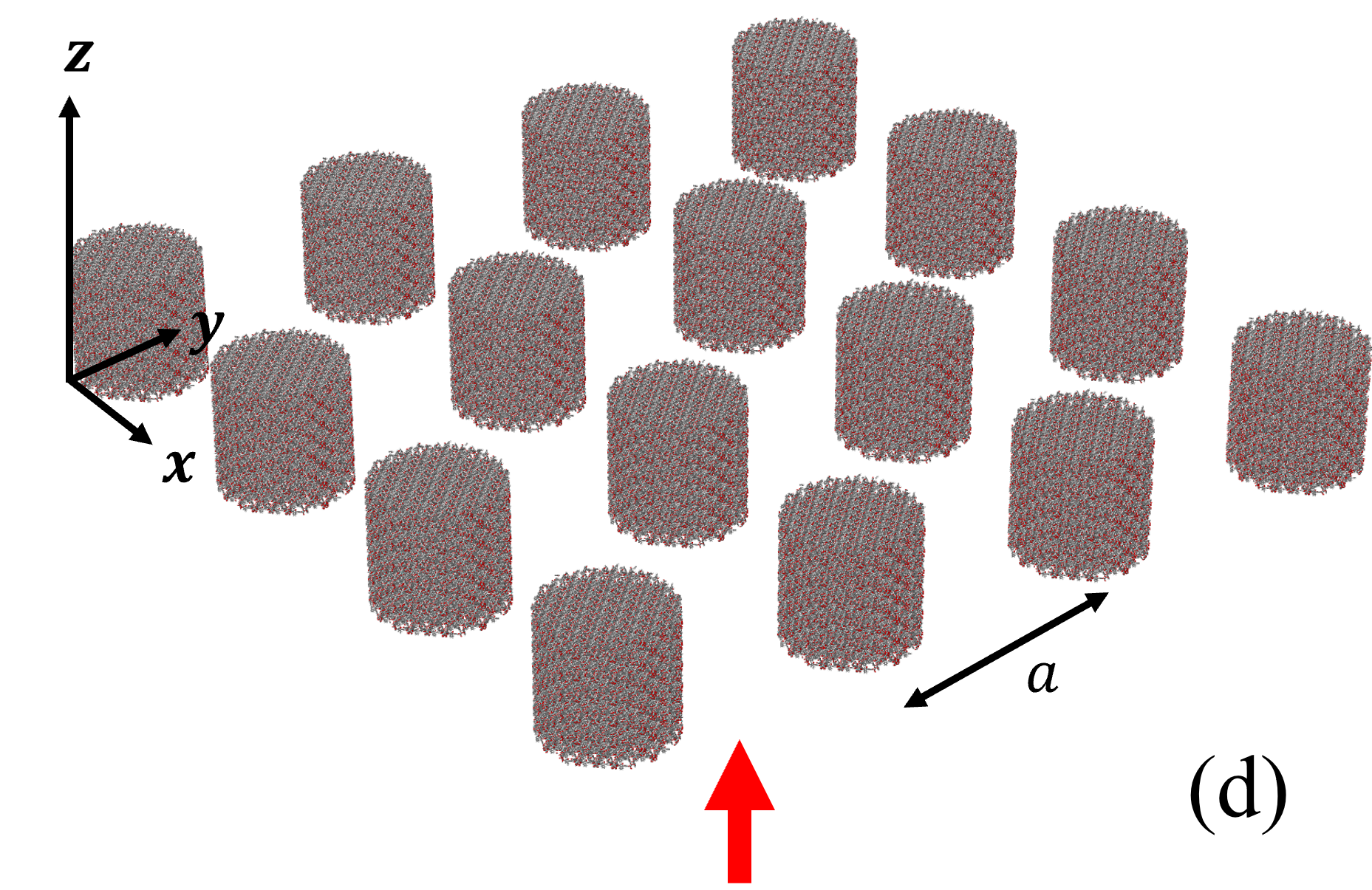}
  }
	\caption{\textbf{(a)} Schematic setting of the simulation and experiment of the MOF thin film on a glass substrate. \textbf{(b)} Schematic setting of the MOF thin film in a Fabry-Pérot cavity defined by two aluminum mirrors. \textbf{(c)} A single cylinder and \textbf{(d)} a metasurface consisting of a square lattice of cylinders made out of the MOF. For the MOF thin film on the substrate, the glass substrate is treated as infinitely extended into the upper half-space. For the MOF thin film in the cavity, the thickness of the lower mirror is 10\,nm and the thickness of the upper mirror is 30\,nm. The MOF cylinders with radius $r_{\mathrm{cyl}}$ and height $h_{\mathrm{cyl}}$ are placed into a square lattice with lattice constant $a$.
	In the simulations, the MOF thin film, the cavity, and the lattice are sequentially illuminated from below with a left- and right-handed circularly polarized plane wave. The transmission is measured in the half-space above the MOF film or the upper mirror, and the reflection is measured in the half-space below the MOF film or below the lower mirror, respectively. }
    \label{fig:SURMOFSubs}
	\end{figure*}

As illustrated in Figure~\ref{fig:SURMOFSubs}\textbf{(a)}, the MOF thin film is illuminated from below, the transmitted light intensity is computed in the half-space of the glass substrate above the MOF, and the reflected light intensity is computed in the half-space below the MOF. To avoid an influence on the spectra by assuming a specific finite value for the thickness of the substrate, we assume that it fills the entire half-space above the MOF. The dielectric function of the glass substrate is assumed to be non-dispersive and set to $\varepsilon_{\mathrm{r}}=2.25$.
We analyze the absorption and circular dichroism of the chiral UiO-BINOL MOF thin film on the substrate. The absorption is defined as $A_{\pm}=1-T_{\pm}-R_{\pm}$, where $T_{\pm}$ is the transmittance and $R_{\pm}$ is the reflectance of an incident plane wave which is either left-handed circularly polarized ($+$) or right-handed circularly polarized $(-)$. The absorption circular dichroism (ACD) is defined as $\mathrm{ACD}=\frac{A_+-A_-}{2}$. We additionally define the transmission circular dichroism (CD) as 
\begin{align}\label{eq:CDmdeg}
     \mathrm{CD\ (mdeg)}=\arctan\left(\frac{\sqrt{T_-}-\sqrt{T_+}}{\sqrt{T_-}+\sqrt{T_+}}\right)\cdot\frac{180\cdot 1000}{\pi}\mathrm{.}
\end{align}
 The unit ``mdeg'' is very common for spectroscopic measurements of circular dichroism in chemistry and biology; see \cite{Gu2014}, for instance. We consider the absorption and circular dichroism for different wavelengths of the incident wave and different thickness values of the film before we concentrate on specific thickness values. 

In the experiments, the light spot illuminates a fairly large area of the sample in which the crystal has multiple domains that are differently oriented to each other, see Figure~\ref{fig:StructuresXRD}\textbf{(d)}. To compare simulations to experiments, we, therefore, calculate the rotational average of the four block matrices of the T-matrix of the unit cell and then solve the multi-scattering problem. In the Supporting Information, we also discuss the macroscopic response of the slab without calculating the rotational average, which then incorporates all the structural characteristics and the orientation of the unit cell. 

\begin{figure*}
\centering
     \subfloat{
	\includegraphics[width=0.32\textwidth]{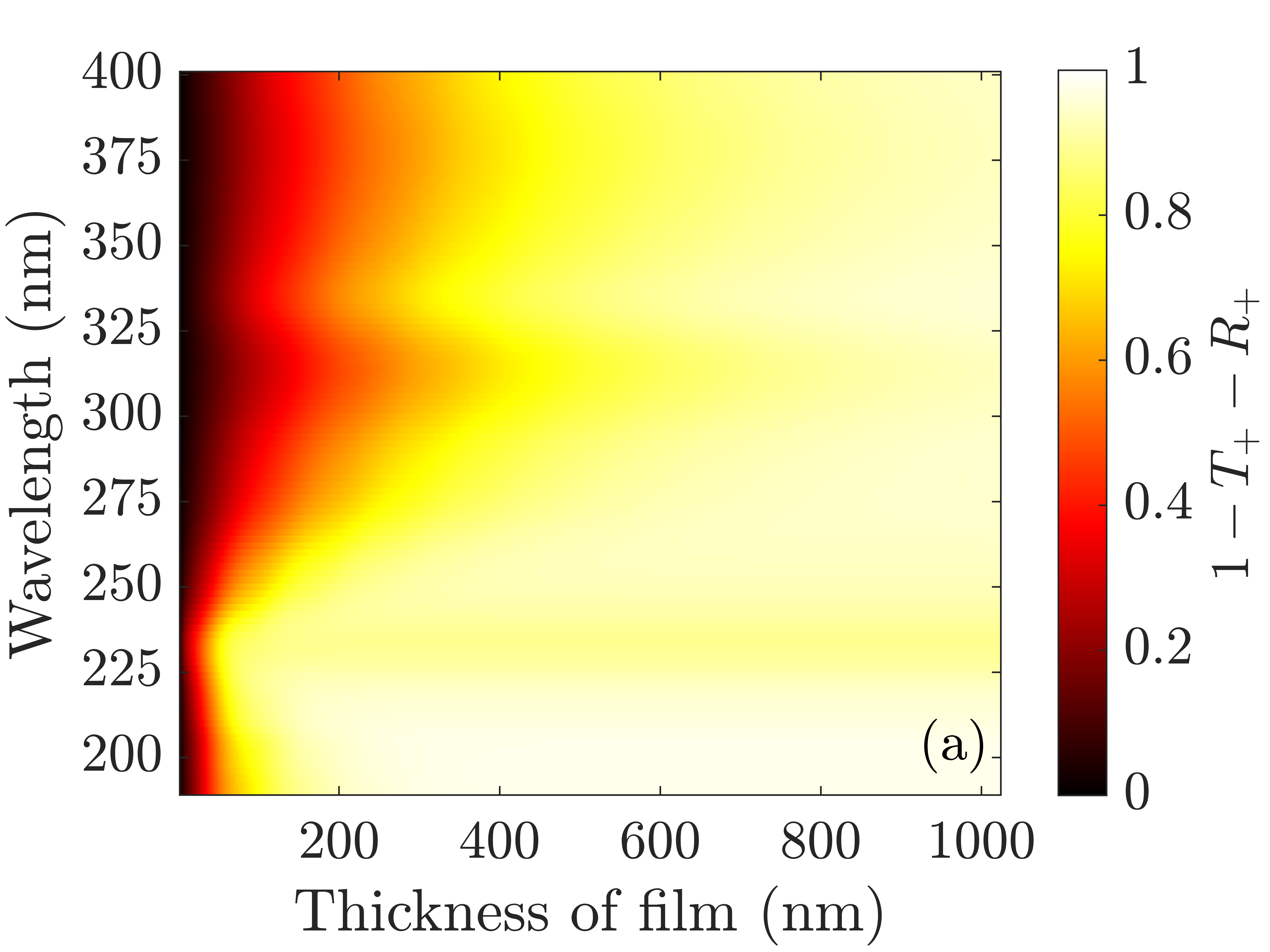}
	}\hspace{0cm}
	\subfloat{
	\includegraphics[width=0.32\textwidth]{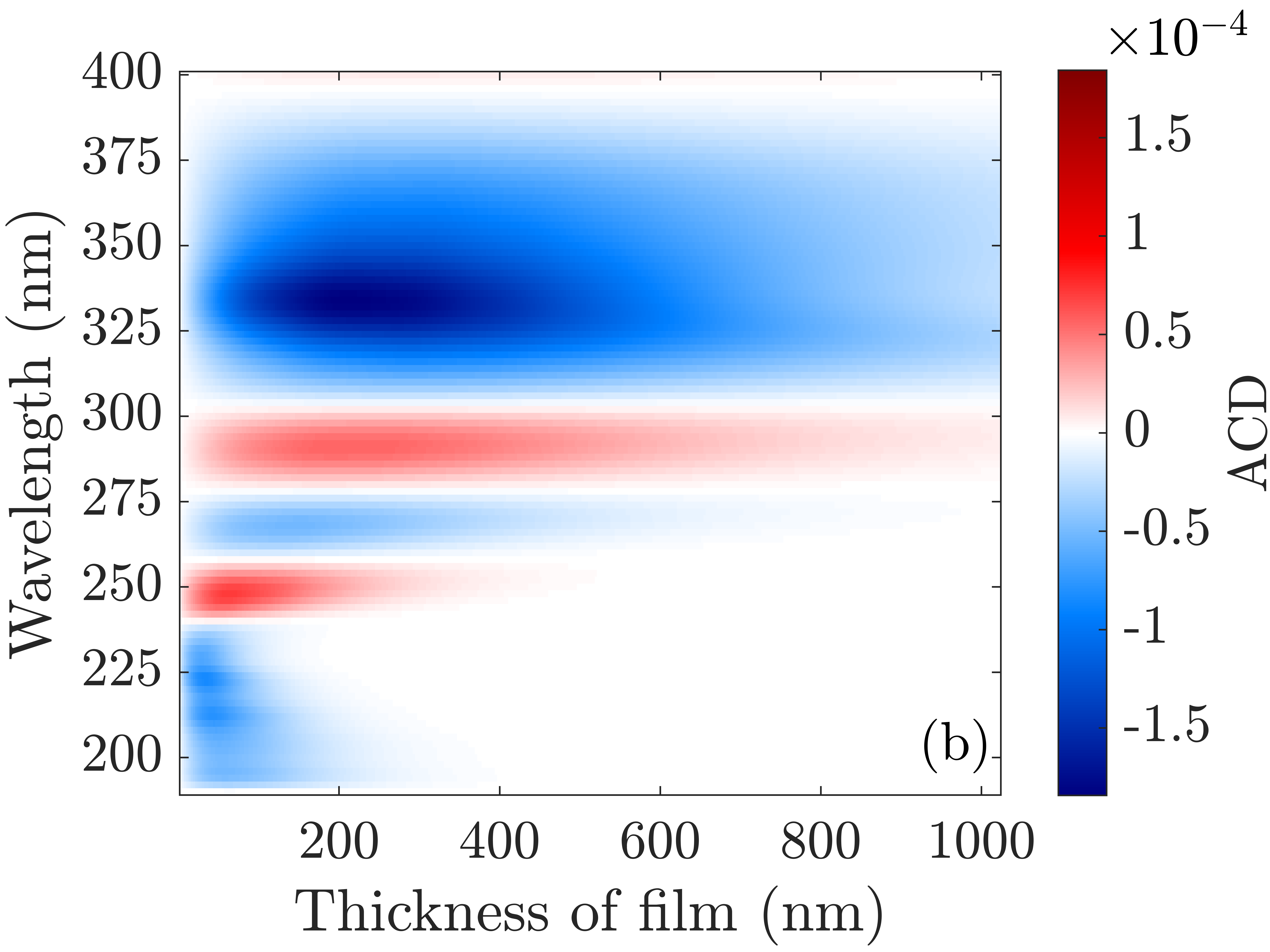}
	}
 \subfloat{
	\includegraphics[width=0.32\textwidth]{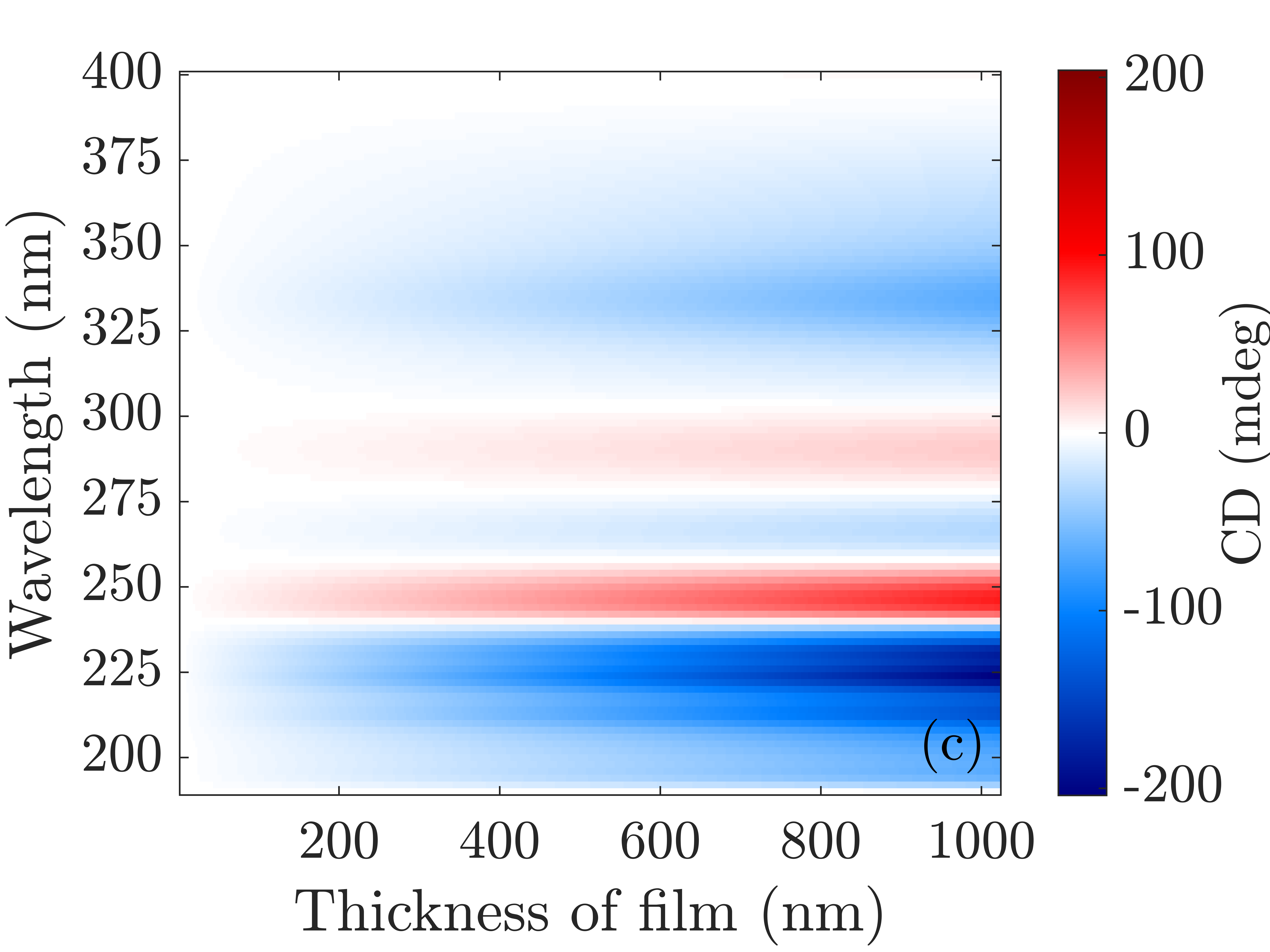}
	}
	\caption{\textbf{(a)} Simulated absorption, \textbf{(b)} ACD, and \textbf{(c)} CD spectra of the UiO-\textit{R}-BINOL MOF as a function of the wavelength and the film thickness. In contrast to the absorption, which is relatively high in the entire parameter range, the ACD and the CD spectra show some distinct regions with alternating signs depending on the wavelength of the incident light.}
    \label{fig:AbsCDRBinolRotAvg}
	\end{figure*}

The ACD of the UiO-\textit{R}-BINOL MOF thin film is depicted in Figure~\ref{fig:AbsCDRBinolRotAvg}\textbf{(a)}, the ACD spectrum is depicted in Figure~\ref{fig:AbsCDRBinolRotAvg}\textbf{(b)}, and the CD spectrum is depicted in Figure~\ref{fig:AbsCDRBinolRotAvg}\textbf{(c)}. While the absorption is relatively high throughout the parameter range, the ACD and CD spectra show distinct regions of positive and negative signs depending on the illumination wavelength. In contrast to the CD spectrum, the ACD spectrum takes relatively small values for short wavelengths and large values of the film thickness. We discuss this difference below by considering all three quantities for specific thicknesses. We show in the Supporting Information that the circular dichroism spectra of the UiO-\textit{S}-BINOL MOF thin film are mirror images of the corresponding spectra of the UiO-\textit{R}-BINOL MOF thin film. Hence, these spectra are not shown here for brevity.

\begin{figure*}
\centering
     \subfloat{
	\includegraphics[width=0.32\textwidth]{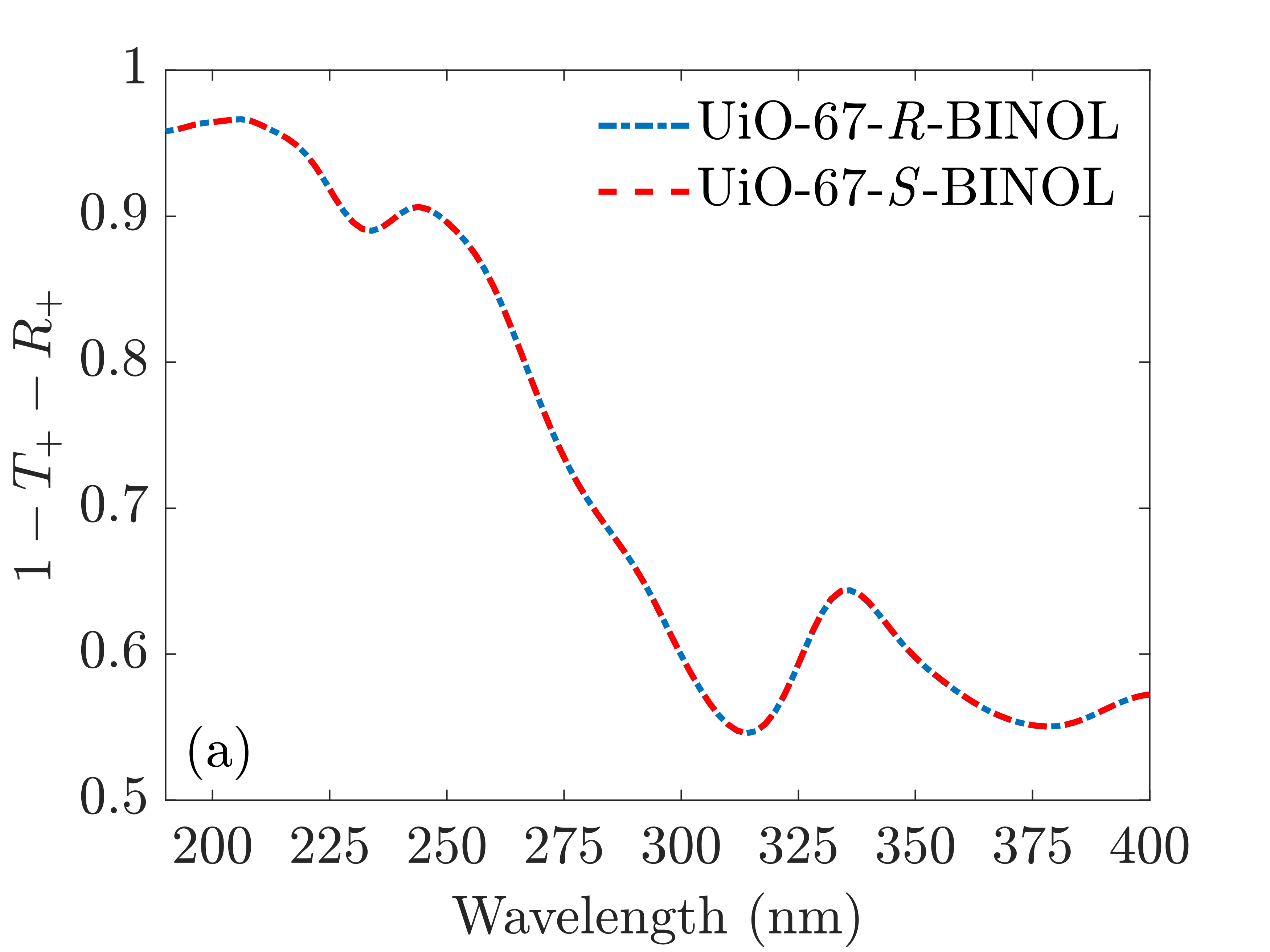}
	}
	\subfloat{
	\includegraphics[width=0.32\textwidth]{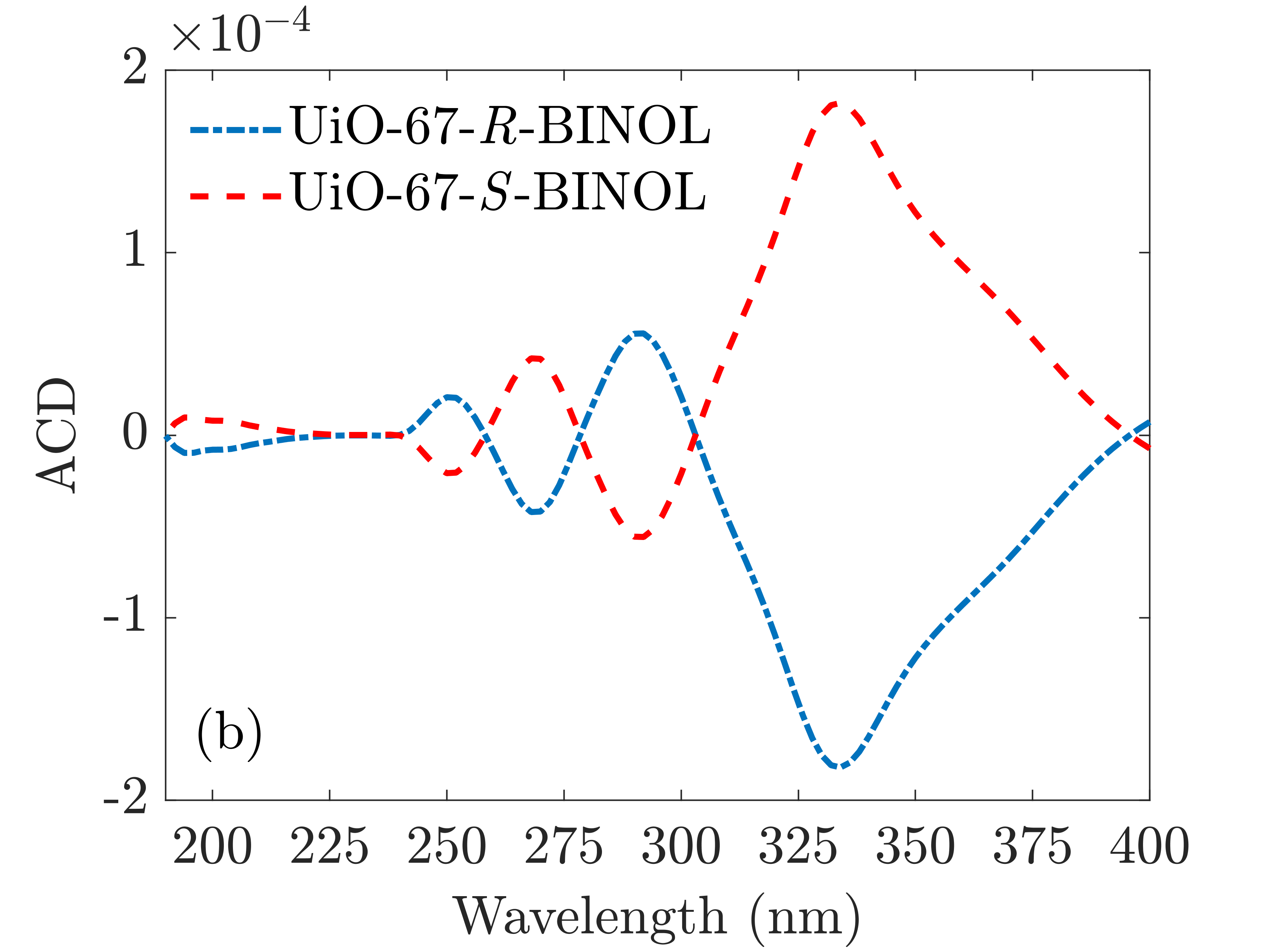}
	}
 \subfloat{
	\includegraphics[width=0.32\textwidth]{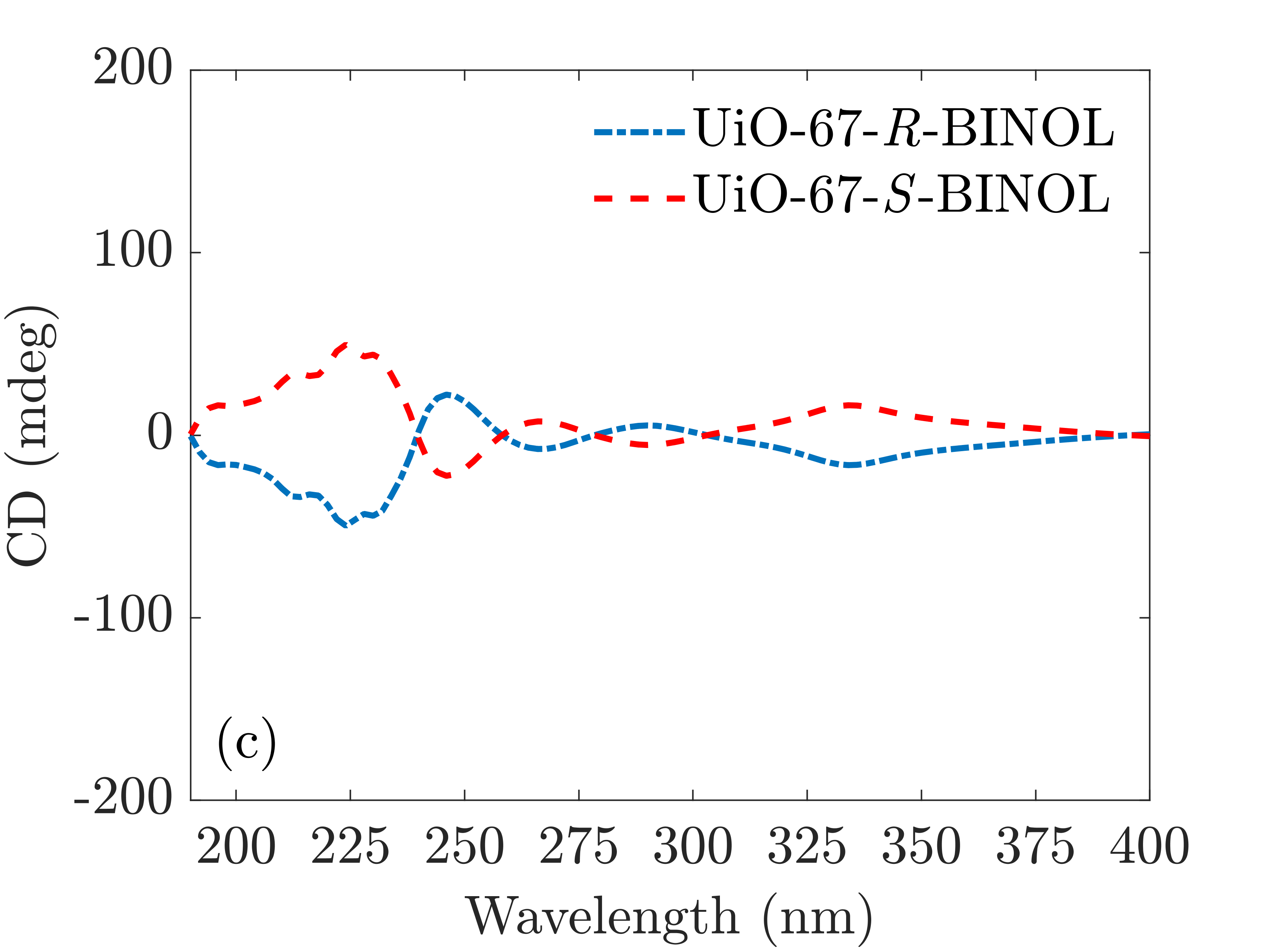}
	}\\
 \subfloat{
	\includegraphics[width=0.32\textwidth]{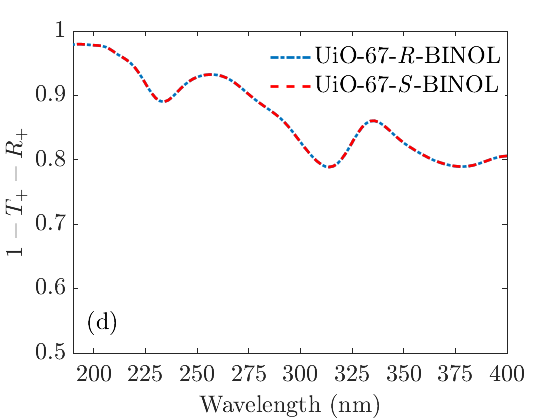}
	}
	\subfloat{
	\includegraphics[width=0.32\textwidth]{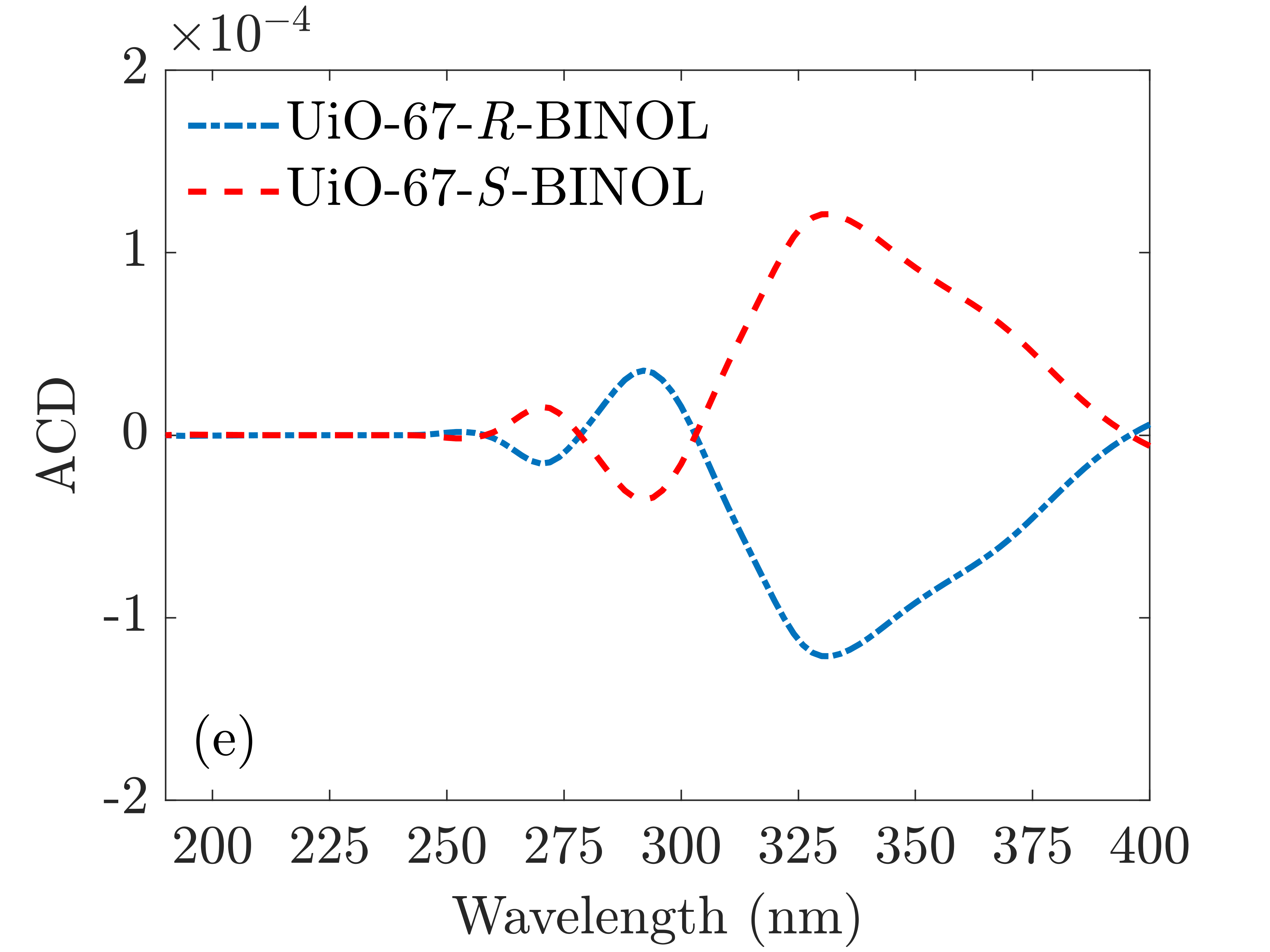}
	}
 \subfloat{
	\includegraphics[width=0.32\textwidth]{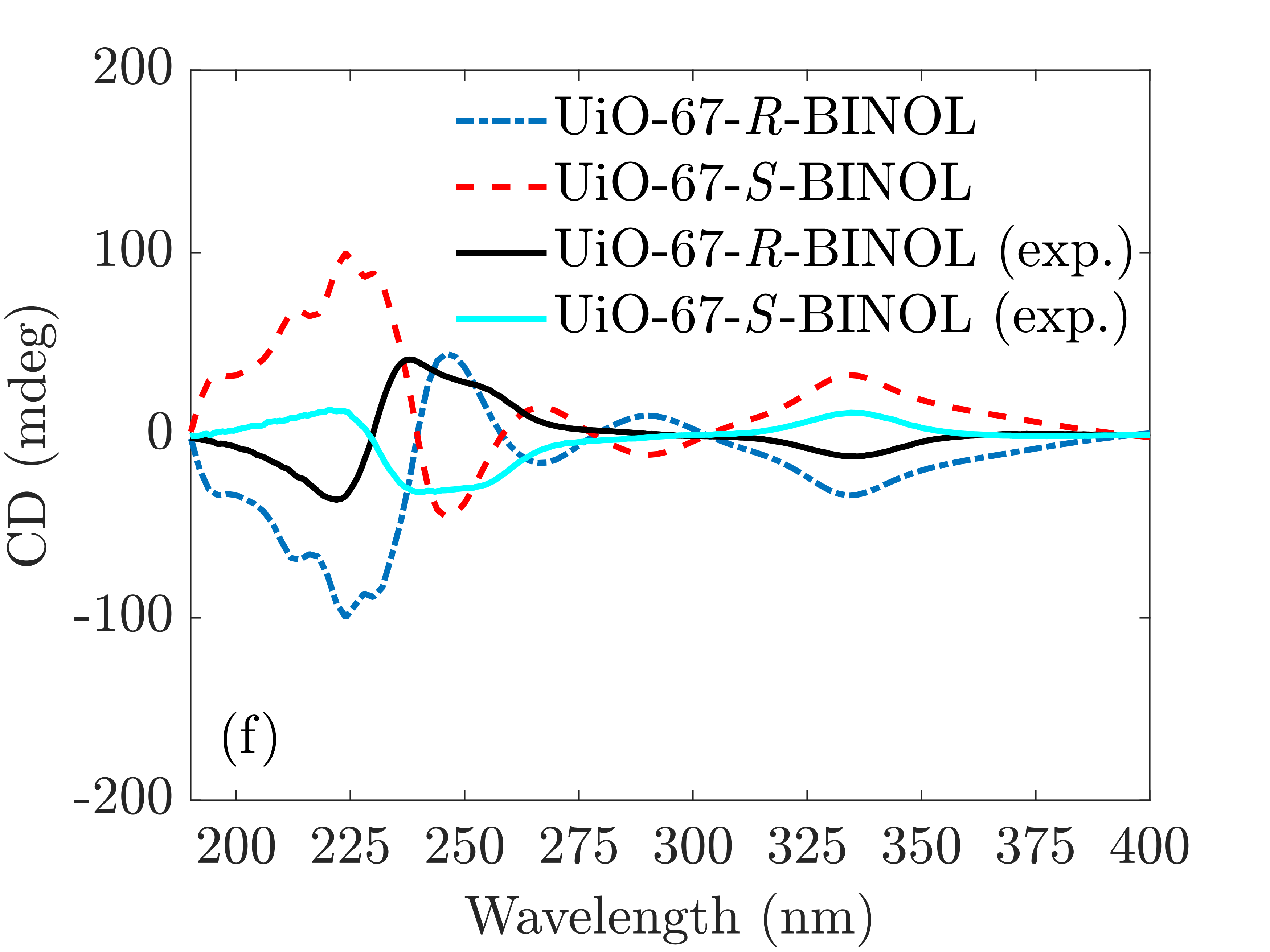}
	}\\

  \subfloat{
	\includegraphics[width=0.32\textwidth]{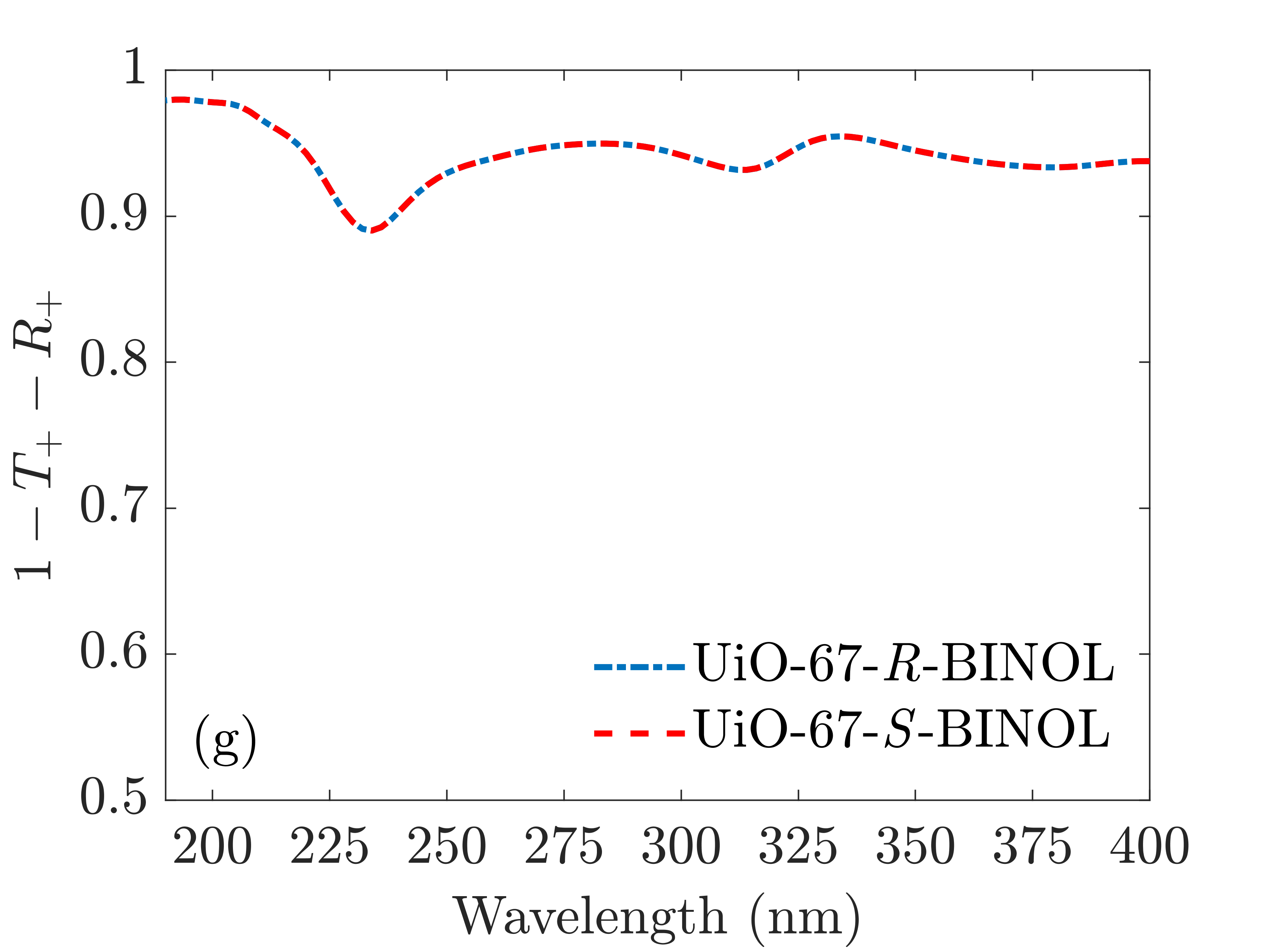}
	}
	\subfloat{
	\includegraphics[width=0.32\textwidth]{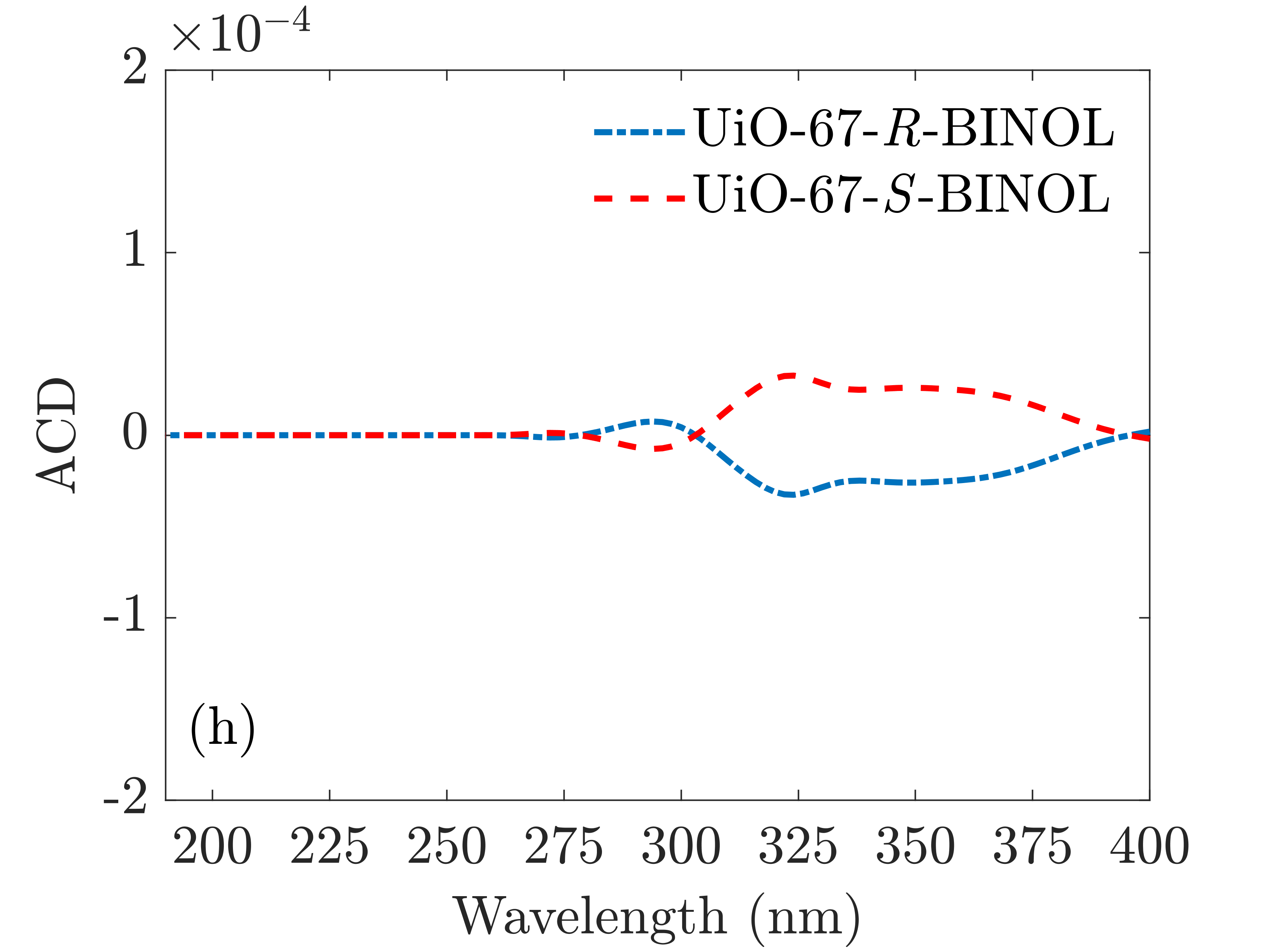}
	}
 \subfloat{
	\includegraphics[width=0.32\textwidth]{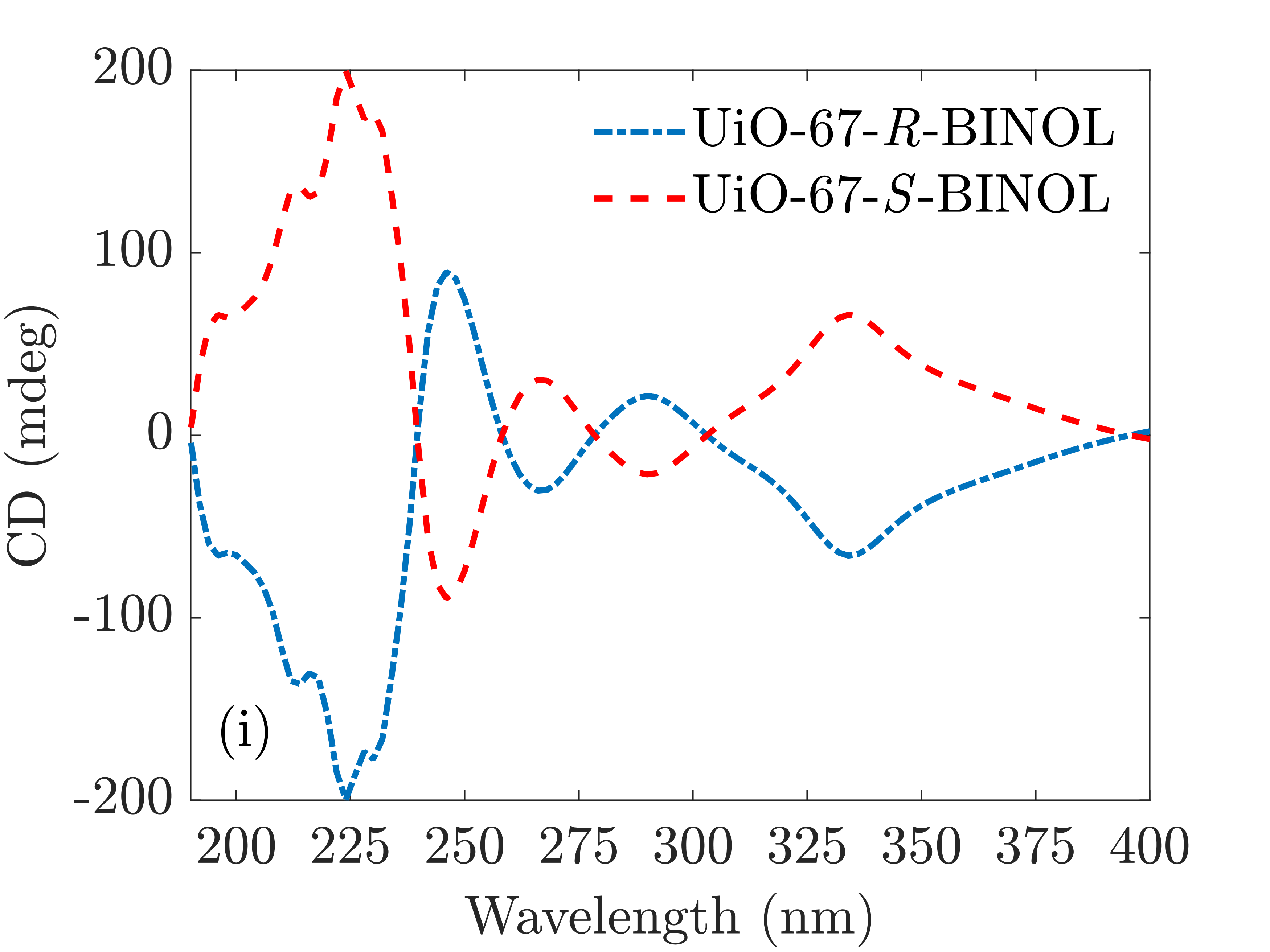}
	}
	\caption{ \textbf{(a)},\textbf{(d)},\textbf{(g)} Simulated absorption,  \textbf{(b)},\textbf{(e)},\textbf{(h)} ACD, and \textbf{(c)},\textbf{(f)},\textbf{(i)} CD spectra for a thickness of the MOF film of 250\,nm \textbf{(a)-(c)}, 500\,nm \textbf{(d)-(f)}, and 1000\,nm \textbf{(g)-(i)}. With increasing thickness (going down the column), the absorption increases. Due to the decreasing signal, the ACD decreases with increasing absorption. The CD, however, increases with growing thickness and absorption as the denominator in Equation~(\ref{eq:CDmdeg}) decreases. Compared to the experimental measurement, the simulated spectra of the CD agree well concerning the resonance frequencies. Concerning the peak intensity, the spectra match relatively well for a thickness of 500\,nm \textbf{(f)}.}
    \label{fig:ThicknessRotAvg}
	\end{figure*}

To analyze the mentioned properties of the circular dichroism spectra, we focus on specific values of the thickness and compare simulated to measured spectra. We choose as values of the thickness 250\,nm, 500\,nm, and 1000\,nm. In Figures~\ref{fig:ThicknessRotAvg}\textbf{(a)}, \textbf{(d)}, and \textbf{(g)}, the absorption of the UiO-\textit{R}-BINOL MOF and UiO-\textit{S}-BINOL MOF thin films are depicted. In Figures~\ref{fig:ThicknessRotAvg}\textbf{(b)}, \textbf{(e)}, and \textbf{(h)} the ACD spectrum, and in Figures~\ref{fig:ThicknessRotAvg}\textbf{(c)}, \textbf{(f)}, and \textbf{(i)} the CD spectrum are shown. 
 
The thickness of the film increases from 250\,nm in the first row to 1000\,nm in the third row in Figure~\ref{fig:ThicknessRotAvg}. The absorption increases with growing film thickness. The ACD simultaneously decreases as the transmission signal becomes smaller for a higher thickness. The CD, however, increases for higher thickness values as the denominator in Equation~(\ref{eq:CDmdeg}) decreases. The thickness of the experimentally analyzed MOFs is in the range between 500\,nm and 1000\,nm. The experimentally measured spectra of the CD are plotted together with the simulations for a thickness of 500\,nm in Figure~\ref{fig:ThicknessRotAvg}\textbf{(f)}. The frequencies with resonances in the CD spectra agree very well, and also the intensities of the peaks match relatively well. The good agreement we observe proves that, on the one hand, the grown MOF thin films are of very good quality due to the agreement with the results of the DFT simulations of the unit cell and with the results of the simulations of the macroscopic samples. Differences between the thickness of the grown film and the simulated film and possible inaccuracies of the measurement setup might cause observed discrepancies in the intensity. For a thickness of 250\,nm of the simulated film, for instance, the agreement between the peak intensities in the simulated and the measured spectra is better. On the other hand, the good agreement supports that our simulation workflow and the used simulation tools are very useful for analyzing and interpreting experimental results. Furthermore, they can be used to design devices based on the given MOF material. The next section is focused on the discussion of specifically tailored devices based on the UiO-67-\textit{R}-BINOL MOF. 

\section{Simulation of Complex Photonic Devices of UiO-BINOL MOF}\label{Sec:Applications}
In this section, we simulate two complex photonic devices incorporating the UiO-67-\textit{R}-BINOL MOF: a Fabry-Pérot cavity consisting of two separated aluminum mirrors filled with the MOF thin film and a metasurface defined by a square lattice of cylinders consisting of the MOF.

\subsection{Optical cavity filled with UiO-67-\textit{R}-BINOL MOF thin film}
\begin{figure*}
\centering
     \subfloat{
	\includegraphics[width=0.32\textwidth]{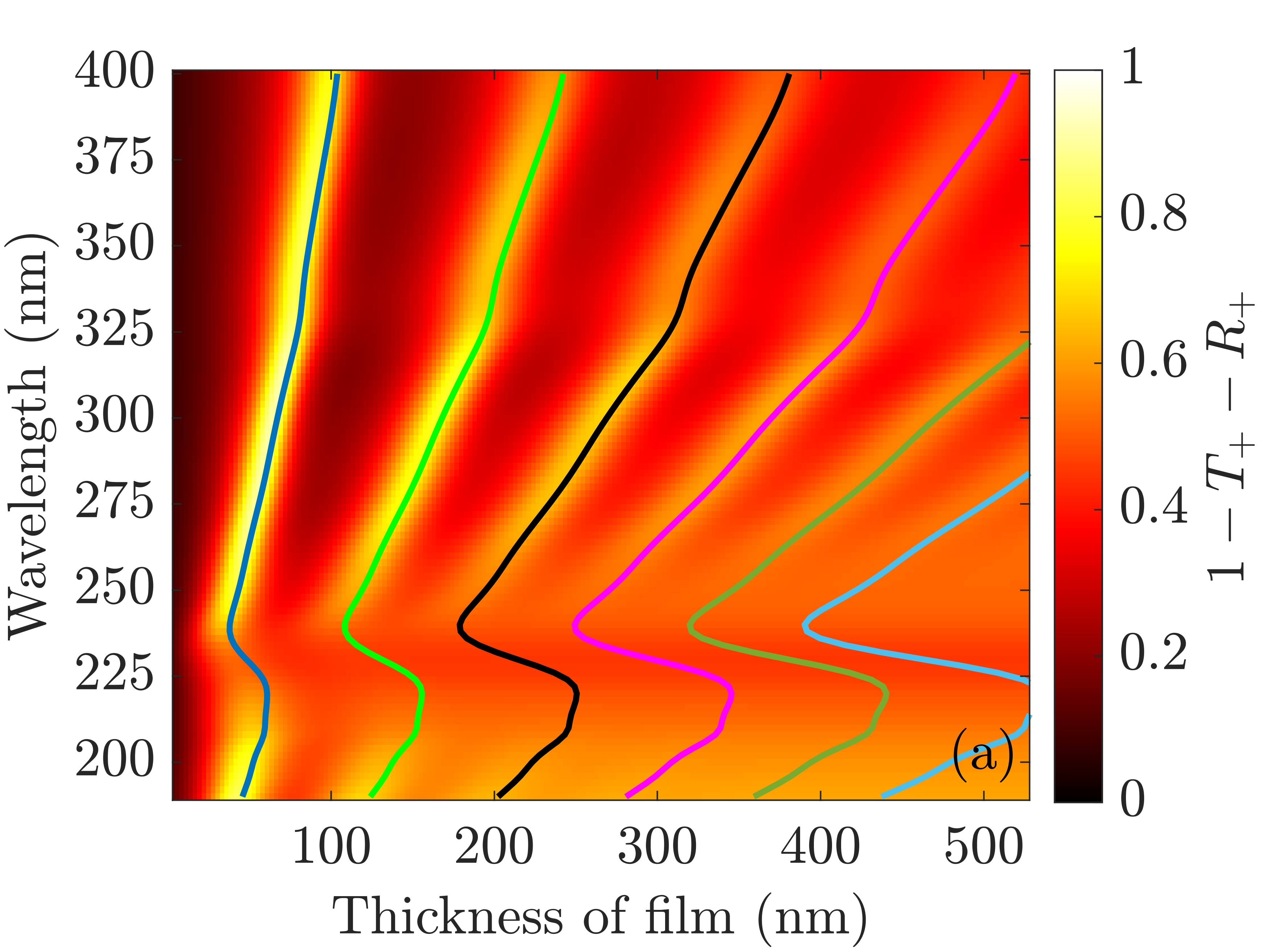}
	}\hspace{0cm}
	\subfloat{
	\includegraphics[width=0.32\textwidth]{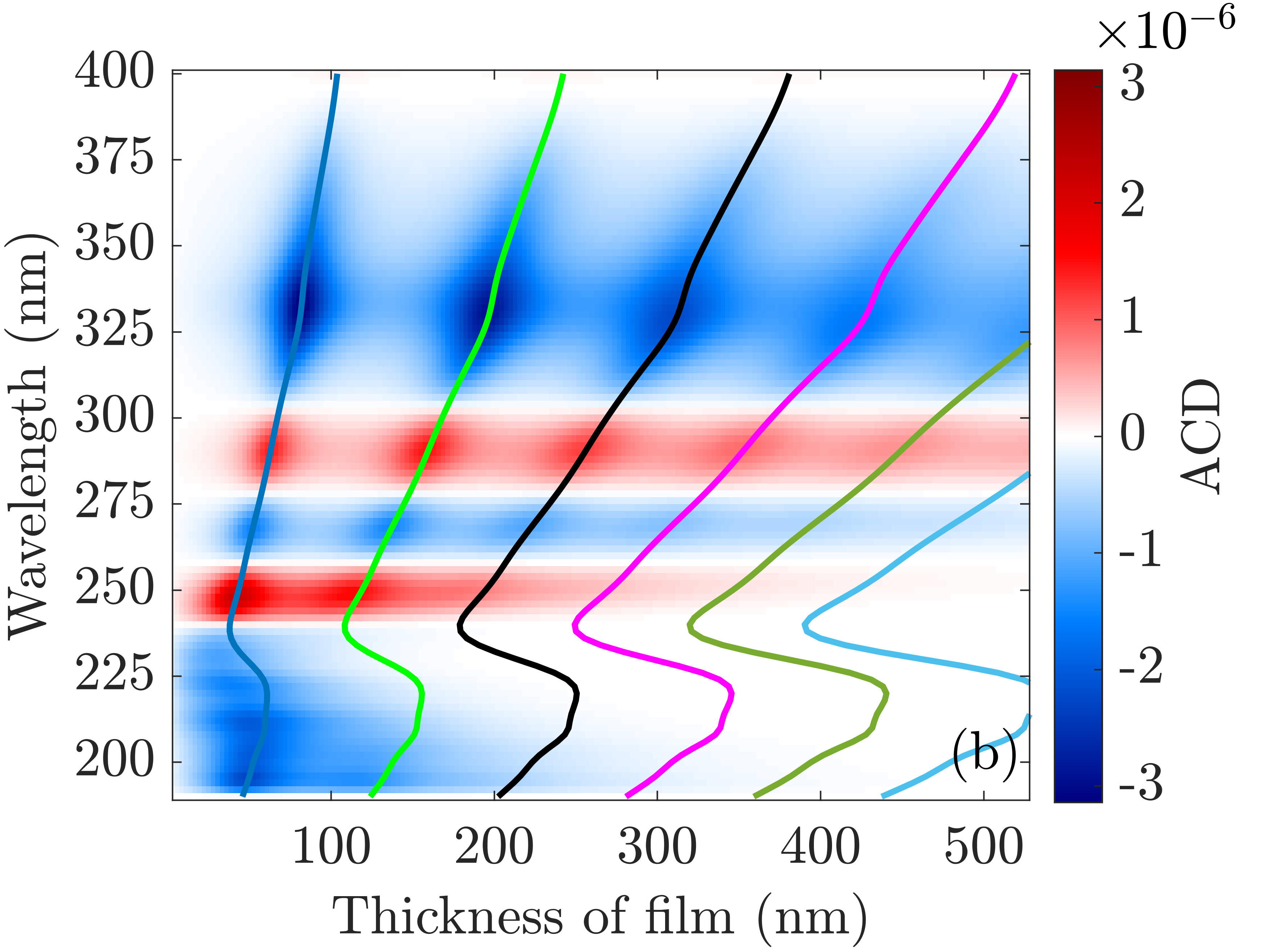}
	}
 \subfloat{
	\includegraphics[width=0.32\textwidth]{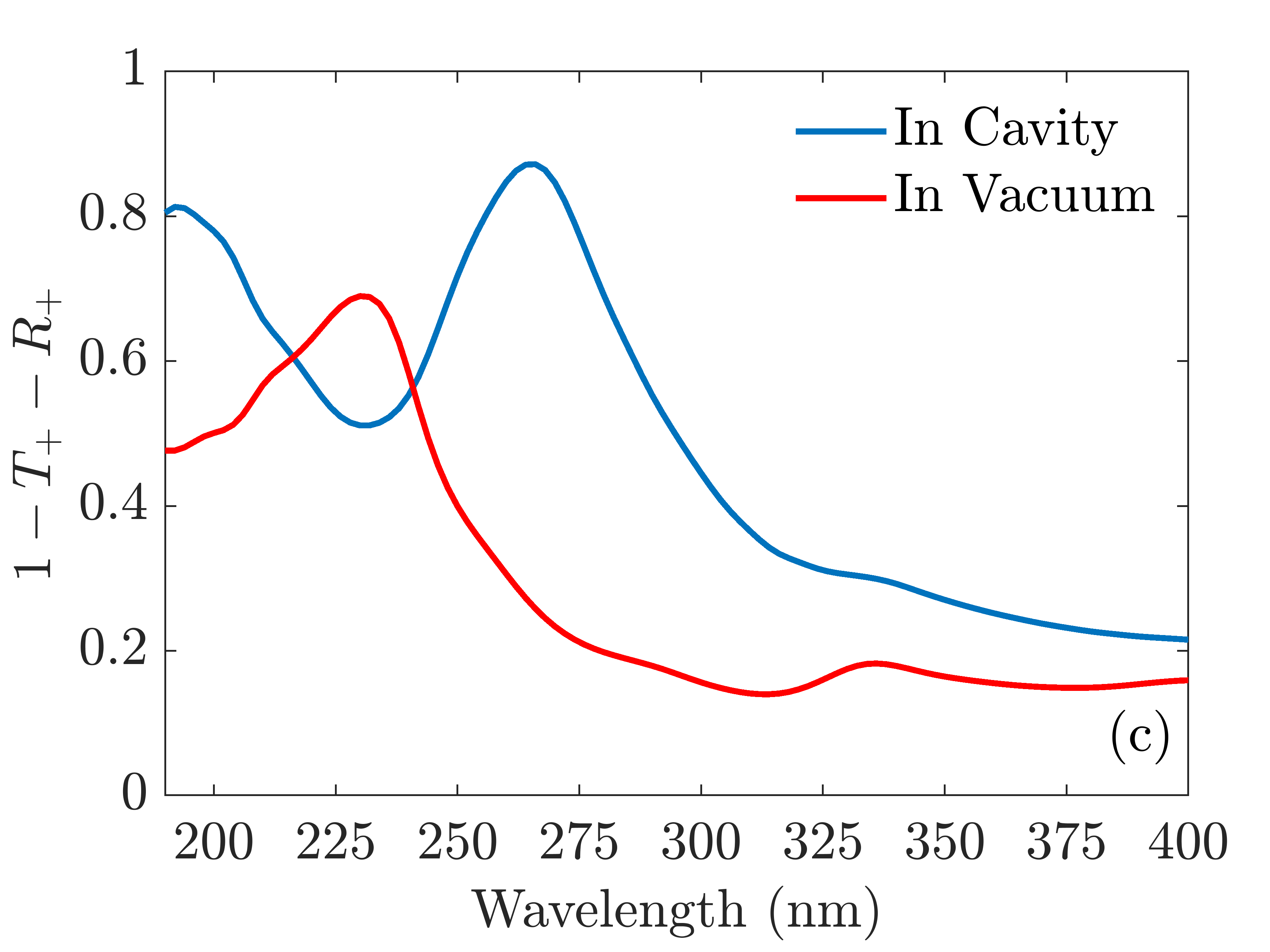}
	}
	\caption{\textbf{(a)} Simulated absorption and \textbf{(b)} ACD spectra depending on the wavelength of the incident light and the MOF film thickness when placed into a Fabry-Pérot cavity made from two thin silver films. \textbf{(c)} Simulated absorption for a fixed thickness of 47\,nm of the MOF film inside the same Fabry-Pérot cavity. The absorption shows regions of large intensity corresponding to cavity modes. The modes are bent around a wavelength of 225\,nm due to the interaction of light and the MOF material. The ACD signal is considerably smaller compared to the MOF thin film on the substrate due to the interaction of the reflected light waves inside the cavity and the higher absorption connected to the reflection of the mirrors. The absorption spectrum in \textbf{(c)} shows the splitting of a cavity mode due to the light-matter coupling.}
    \label{fig:CavityRBinolRotAvg}
	\end{figure*}

We consider a Fabry-Pérot cavity filled with the UiO-67-\textit{R}-BINOL MOF thin film; see Figure~\ref{fig:SURMOFSubs}\textbf{(b)}. The cavity consists of two separated thin aluminum films that act as mirrors. For the optical simulations, the material parameters from aluminum are taken from \cite{doi:10.1021/ph5004237}. The bottom mirror has a thickness of 10\,nm, and the top mirror has a thickness of 30\,nm. The cavity is illuminated with a left-handed circularly polarized plane wave from the bottom propagating in $+z$-direction.  
In the optical simulations of the MOF thin film, we consider again the rotationally averaged T-matrix of the unit cell of the MOF to reflect the fact that the thin film is not mono-crystalline. Instead, it consists of domains with different orientations. In Figure~\ref{fig:CavityRBinolRotAvg}\textbf{(a)}, the absorption spectrum is shown for different wavelengths of the incident light and different values of the thickness of the MOF thin film. The thickness of the film is restricted to values below 500\,nm as the high absorption for larger thicknesses suppresses the observation of finer details. Compared to Figure~\ref{fig:AbsCDRBinolRotAvg}\textbf{(a)}, one observes narrow regions of high absorption which are bent close to the resonances of the MOF. Indeed, these lines of high absorption can be traced to the excitation of standing wave resonances in the Fabry-Pérot cavity filled with a dispersive material \cite{SURMOFCavity}. The modes of the cavity, depicted as colored lines in Figure~\ref{fig:CavityRBinolRotAvg}\textbf{(a)}, are calculated based on the effective refractive index of the MOF, which is computed with the homogenization method from \cite{HomogenEffTMat}. For the calculation of the cavity modes, the chirality of the MOF is neglected. 

In Figure~\ref{fig:CavityRBinolRotAvg}\textbf{(b)}, the absorption spectrum is depicted. In comparison to Figure~\ref{fig:AbsCDRBinolRotAvg}\textbf{(b)}, the values of the ACD are considerably smaller for the MOF thin film in the cavity. On the one hand, this can be related to the fact that for building a cavity, planar mirrors are unsuitable for enhancing the circular dichroism. The circular polarization of the incident light flips the handedness upon each reflection inside the cavity. Therefore, the circular dichroism signal does not add up but rather tends to cancel \cite{PhysRevLett.124.033201}. On the other hand, a high number of reflections leads to larger absorption, which decreases the absorption circular dichroism similar to Figure~\ref{fig:ThicknessRotAvg}. Therefore, the cavity modes decrease the signal compared to the configuration with the glass substrate. However, comparing the values of the ACD at different thickness values, the signal is larger in those parameter regimes where the cavity supports a mode. This can be related to the fact that outside those parameter regimes, destructive interference between the light waves occurs, and the net ACD vanishes. The CD spectrum is not shown here explicitly. The spectrum does not differ significantly from the spectrum of the MOF thin film on the substrate. That is because the normalization of the CD by the transmitted power suppresses the absorption influence on the signal so that a larger number of reflections does not change the quantity.

In Refs.~\cite{D0SC02436H,SURMOFCavity}, the interaction between a cavity made from silver mirrors and a Zn-SiPc-SURMOF-2 has been discussed experimentally and numerically, respectively. In \cite{SURMOFCavity}, it is analyzed that for fixed values of the thickness of the MOF film, the two observed branches of high absorption are not belonging to the same split cavity mode but to two different ones. This is related to the strong dispersion of the refractive index of the SURMOF. For low values of the thickness of the UiO-\textit{R}-BINOL-MOF film, the observed split branches in Figure~\ref{fig:CavityRBinolRotAvg}\textbf{(a)} belong to the same cavity mode and not to two different ones. In Figure~\ref{fig:CavityRBinolRotAvg}\textbf{(c)}, the absorption is shown for a fixed thickness of the UiO-\textit{R}-BINOL-MOF film of 47\,nm in the aluminum cavity. The absorption of the filled cavity shows two peaks separated by a dip at a wavelength where the absorption of the MOF thin film in a vacuum shows a peak. This indicates a coupling between the light and the MOF, which is commonly referred to as Rabi splitting \cite{https://doi.org/10.1002/cphc.201200734,doi:10.1021/acs.jpclett.5b02512,Perez-Gonzalez_2014}. With respect to Ref.~\cite{D0SC02436H}, we introduce an additional degree of freedom to the split energy states, as their absorption depends on the polarization handedness of the incident light causing slight differences in the light-matter interaction. This can potentially lead to a larger number of applications. 

\subsection{Metasurface made from UiO-67-\textit{R}-BINOL MOF cylinders}
The possibility of enhancing the circular dichroism of chiral objects by the strong near-fields of plasmonic structures or high-refractive-index dielectric materials as silicon cylinders is well-known \cite{https://doi.org/10.1002/adfm.202204095,doi:10.1021/nl3012787,PhysRevX.2.031010,doi:10.1021/acsphotonics.7b01071,doi:10.1021/acsphotonics.8b01454,doi:10.1021/acsphotonics.8b01365,doi:10.1021/acsphotonics.0c00611,doi:10.1021/acsphotonics.9b01352,doi:10.1063/5.0118263,Zong:22}. A central quantity in the discussion of this enhancement is the optical chirality density defined as \cite{PhysRevLett.104.163901}
\begin{align}\label{eq:OptChiralDens}
C(\bm{r},\omega)=-\frac{\varepsilon_{\mathrm{h}}\omega}{2}\mathrm{Im}\{\bm{E}^{\dagger}(\bm{r},\omega)\bm{B}(\bm{r},\omega)\}
\end{align}
at a position $\bm{r}$ of a molecule in the surrounding medium and at the frequency $\omega$.
The optical chirality density is a measure for the difference of the intensities of left- and right-handed circularly polarized fields; see Equation~(4) in \cite{doi:10.1021/acsphotonics.8b01454}. The enhancement of the circular dichroism of chiral molecules placed on an achiral nanostructure is proportional to the volume-averaged enhancement of the optical chirality density by the structure compared to the optical chirality density in the embedding medium of the molecules without the nanostructure, see Equation~(7) in \cite{doi:10.1021/acsphotonics.8b01454}. Especially if the electric and magnetic dipole resonances of the nanostructure overlap, a high enhancement can be achieved \cite{doi:10.1021/acsphotonics.8b01454,doi:10.1021/acsphotonics.8b01365,doi:10.1021/acsphotonics.9b01352}. In this context, also lattice resonances, which collectively couple multipole moments, are of significant importance \cite{doi:10.1021/acsphotonics.0c00611,doi:10.1063/5.0118263,Zong:22}. 

In this subsection, we consider the enhancement of the circular dichroism signal of the UiO-67-\textit{R}-BINOL MOF thin film at a wavelength of 334\,nm by designing a metasurface defined by a square lattice of cylinders that consist of the MOF material, see Figures~\ref{fig:SURMOFSubs}\textbf{(c)} and \textbf{(d)}. That means that the structured molecular material enhances its signal itself. The wavelength of 334\,nm was chosen because, close to it, the ACD and CD spectra show a maximum, see Figure~\ref{fig:ThicknessRotAvg}. It is emphasized that the MOF material is of far more complexity than silicon, for instance. The main advantage of our multiscale ansatz is the representation of this complexity by the T-matrix. From the rotationally averaged T-matrix, the effective material parameters can be extracted that describe the optical properties of the material at the macroscopic scale \cite{HomogenEffTMat}. The material properties that matter to the problem at hand are the permittivity and chirality. Both effective quantities are shown in the Supporting Information in Figure~S5. The effective parameters can then be considered in a macroscopic full-wave Maxwell solver to explore the optical properties of a structured MOF. Initially, we consider here COMSOL Multiphysics which solves Maxwell's equations using the finite-element method \cite{Comsol}. 

\begin{figure*}
\centering
     \subfloat{
	\includegraphics[width=0.5\textwidth]{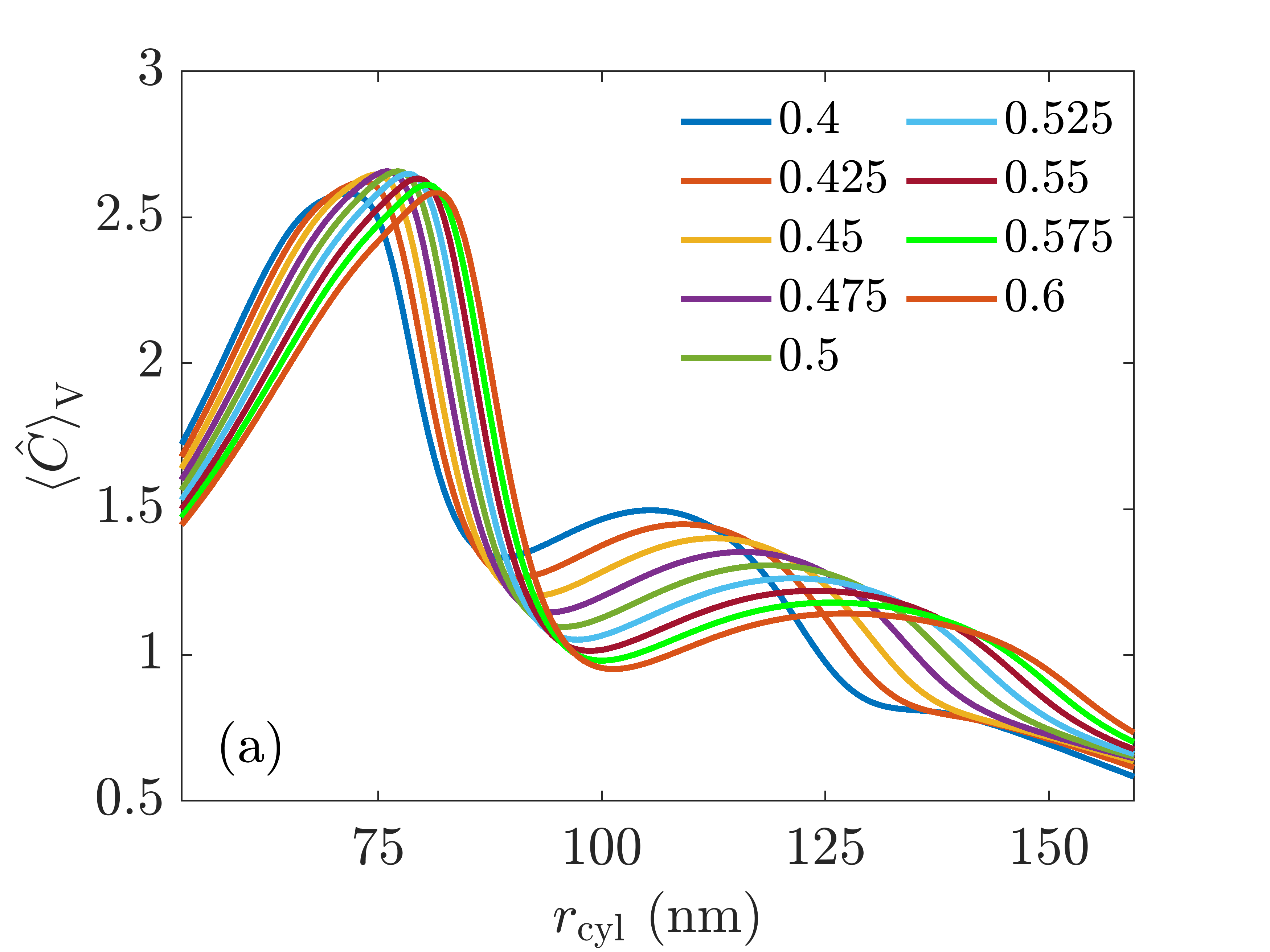}
	}%\hspace{0cm}
	\subfloat{
	\includegraphics[width=0.5\textwidth]{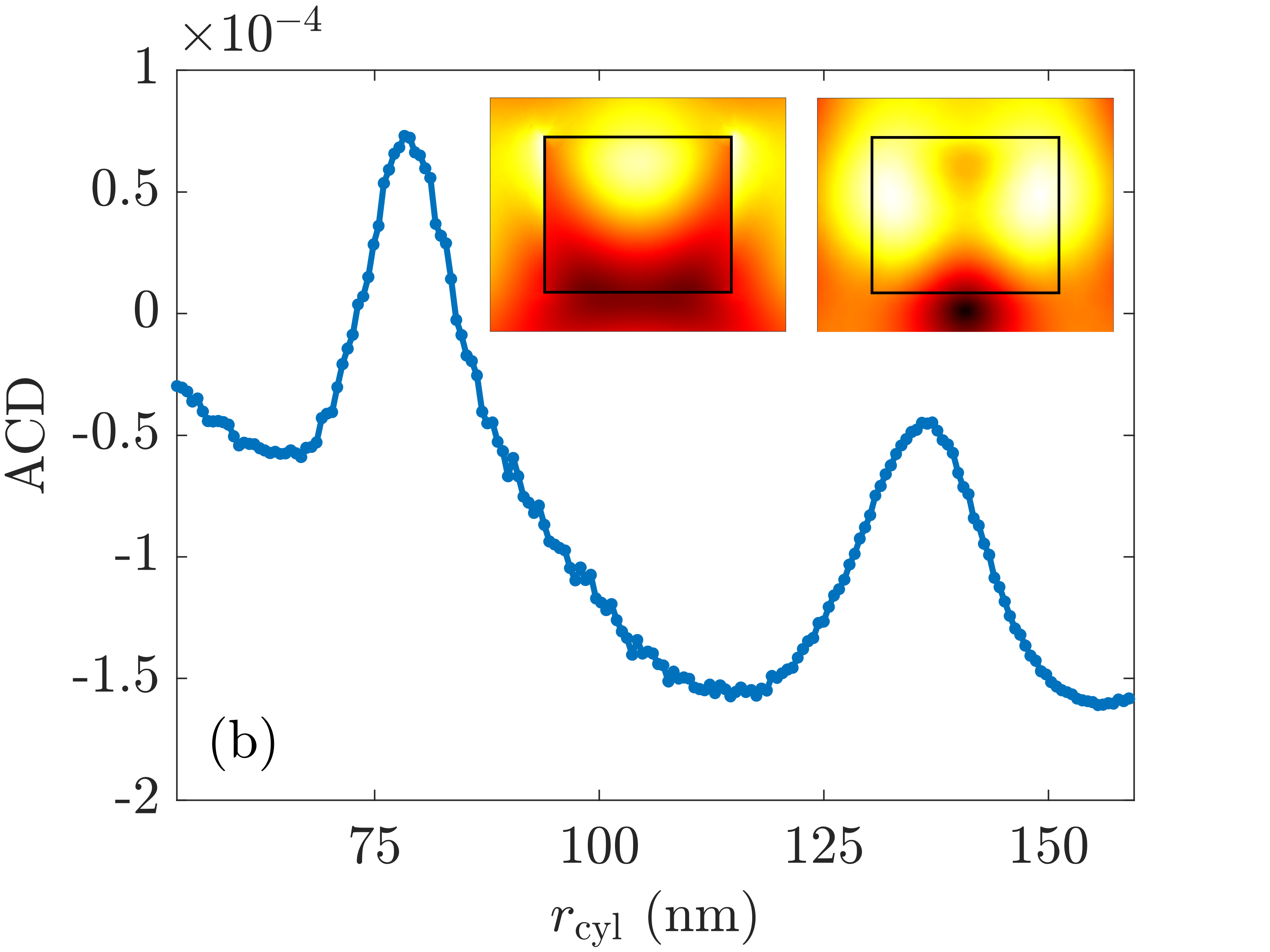}
	}\\
 \subfloat{
	\includegraphics[width=0.5\textwidth]{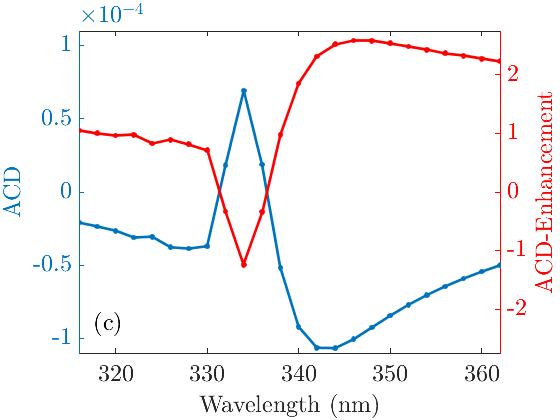}
	}%\hspace{0cm}
 \subfloat{
	\includegraphics[width=0.5\textwidth]{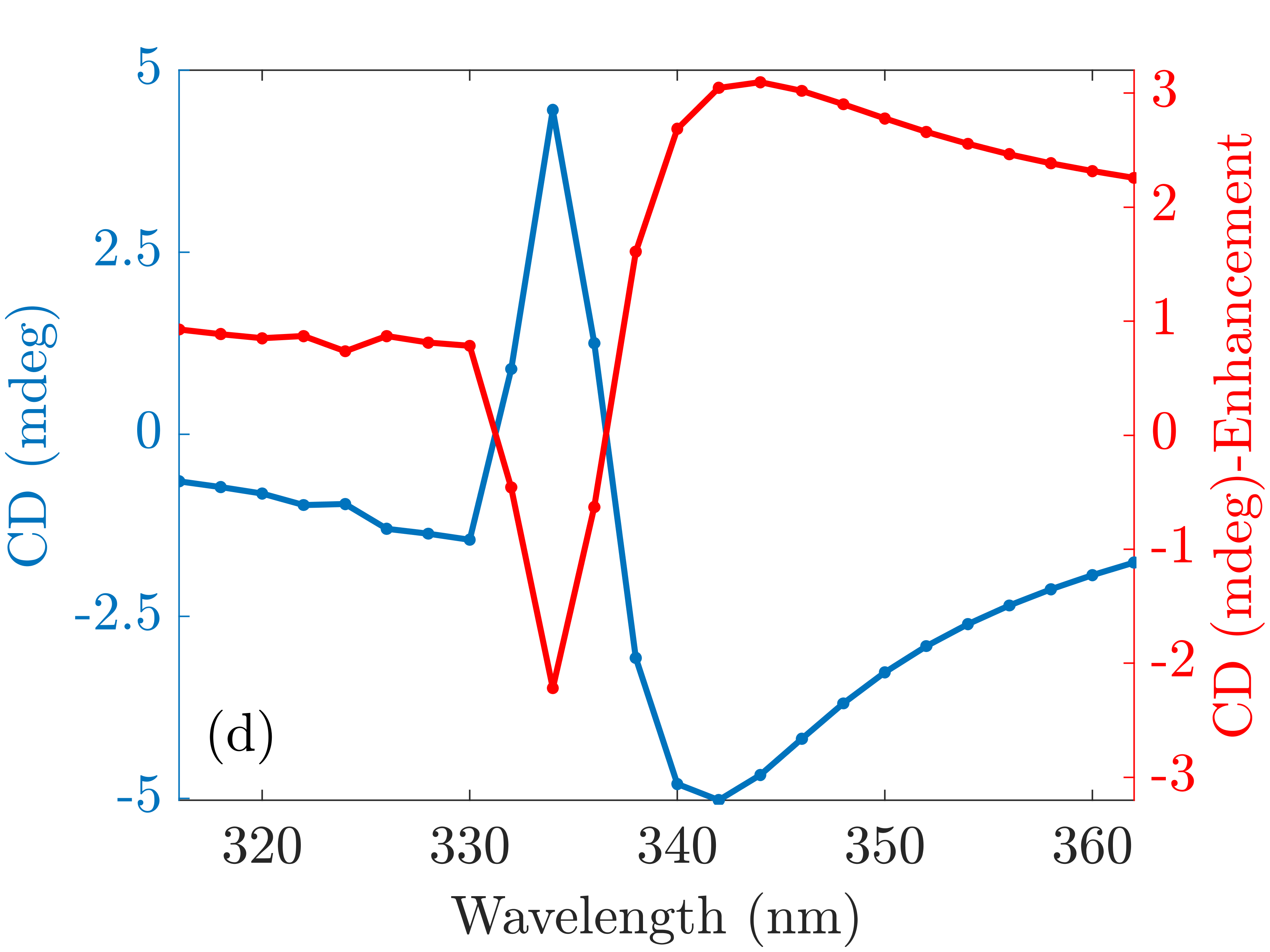}
	}
	\caption{\textbf{(a)} Simulated volume averaged enhancement of the optical chirality density of a MOF cylinder inside a square lattice with lattice constant $a=325.65\,\mathrm{nm}$ for different aspect ratios and radii. \textbf{(b)} Simulated ACD spectrum of the same square lattice of cylinders with an aspect ratio of 0.475 and different radii. The inset images show the absolute values of the electric (left) and magnetic (right) fields inside and close to the cylinder in the lattice for a radius of $r_{\mathrm{cyl}}=78.91\,\mathrm{nm}$. At this radius, the ACD spectrum shows a peak. The fields are normalized by the respective fields of the incident left-circularly polarized wave in the host medium. \textbf{(c)} Simulated wavelength dependent ACD and \textbf{(d)} CD spectra for a cylinder with a radius of $r_{\mathrm{cyl}}=78.91\,\mathrm{nm}$, an aspect ratio of 0.475, and a lattice constant of $a=325.65\,\mathrm{nm}$. The near-field enhancement of the optical chirality density leads to an enhancement of the circular dichroism around the design wavelength of 334\,nm in the shape of a Fano resonance, in comparison to a slab of the same volume in one unit cell of the square lattice.}
    \label{fig:CDRBinolCyl}
	\end{figure*}

In the first step, the optical chirality density for incident left-handed circularly polarized light at the above-mentioned design wavelength is calculated for a metasurface made from MOFs structured into cylinders. We systematically change the cylinder radius and select multiple aspect ratios. The aspect ratio is defined as the ratio between the cylinder's radius and its height. The purpose is to find suitable geometry parameters where the optical chirality density is enhanced. For this specific study, the chirality of the cylinders is neglected as for the case of achiral nanoplatforms surrounded by chiral molecules, see \cite{doi:10.1021/acsphotonics.8b01454,doi:10.1021/acsphotonics.8b01365,doi:10.1021/acsphotonics.9b01352}, for instance. The volume-averaged enhancement of the optical chirality density across the volume occupied by the molecules is, in the case of achiral nanoplatforms, directly proportional to the enhancement of the circular dichroism. Figure~\ref{fig:CDRBinolCyl}\textbf{(a)} shows the volume-averaged enhancement of the optical chirality density with respect to the optical chirality density in vacuum for a lattice constant of $a=325.65\,\mathrm{nm}$. The aspect ratio of $r_{\mathrm{cyl}}/h_{\mathrm{cyl}}=0.475$ is chosen in our further research. In this configuration, the enhancement of the optical chirality density reaches a value of 2.7. As we show below, by choosing a radius close to this peak of the enhancement of the optical chirality density, one observes, for the chosen aspect ratio, resonances of the lattice multipole moments at the design wavelength. These multipole resonances cause strong fields inside the structure.

In the next step, the intrinsic chirality of the MOF is entirely considered in the simulation, and the ACD is calculated. The resulting spectrum is shown for different radii in Figure~\ref{fig:CDRBinolCyl}\textbf{(b)}. The two peaks of the ACD spectrum at $78.91\,\mathrm{nm}$ and $114.61\,\mathrm{nm}$ are related to the enhancement of the optical chirality density caused by strong fields inside the structure, see Figure~\ref{fig:CDRBinolCyl}\textbf{(a)}. For further considerations, we choose the radius $r_{\mathrm{cyl}}=78.91\,\mathrm{nm}$, for which the ACD spectrum shows a peak very close to the most significant peak of the enhancement of the optical chirality density in \textbf{(a)} indicating that high near-fields cause the maximum of the circular dichroism. These near-fields are visualized in the inset of Figure~\ref{fig:CDRBinolCyl}\textbf{(b)}. Below, we demonstrate that for this radius, the structure shows lattice multipole resonances at the design wavelength. In the final step, the wavelength of the incident light is varied in the range around the design wavelength of 334\,nm, taking into account the dispersion of the effective parameters of the MOF material of the cylinders. The ACD and the CD spectra are shown in Figures~\ref{fig:CDRBinolCyl}\textbf{(c)} and \textbf{(d)} with their respective enhancement compared to a slab of the same volume in one unit cell of the square lattice. We observe peaks of the ACD and CD spectra at the design wavelength 334\,nm of $6.94\cdot 10^{-5}$ and 4.45\,mdeg, respectively. The peaks of the respective enhancements at 334\,nm are $-1.24$ and $-2.22$. 

In the following, we analyze the character of the resonance at 334\,nm in the spectra. For that purpose, the T-matrix of the single MOF cylinder is calculated with the finite-element-based program JCMsuite \cite{JCM}. From that, we can also calculate the effective T-matrix of the cylinder inside the two-dimensional lattice at normal incidence. Please note, both T-matrices are different, as the optical response of the individual cylinder gets renormalized (sometimes also called dressed) by the interaction with all the other cylinders forming the lattice. This modifies its optical properties, which needs to be fully considered in the discussion. The effective T-matrix at normal incidence is defined in Equation~(\ref{eq:Teff2D}) for $\bm{k}_{\parallel}=\bm{0}$, and it describes the scattering from a cylinder at the origin of the lattice incorporating the scattering off the other cylinders. 

\begin{figure*}
\centering
     \subfloat{
	\includegraphics[width=0.45\textwidth]{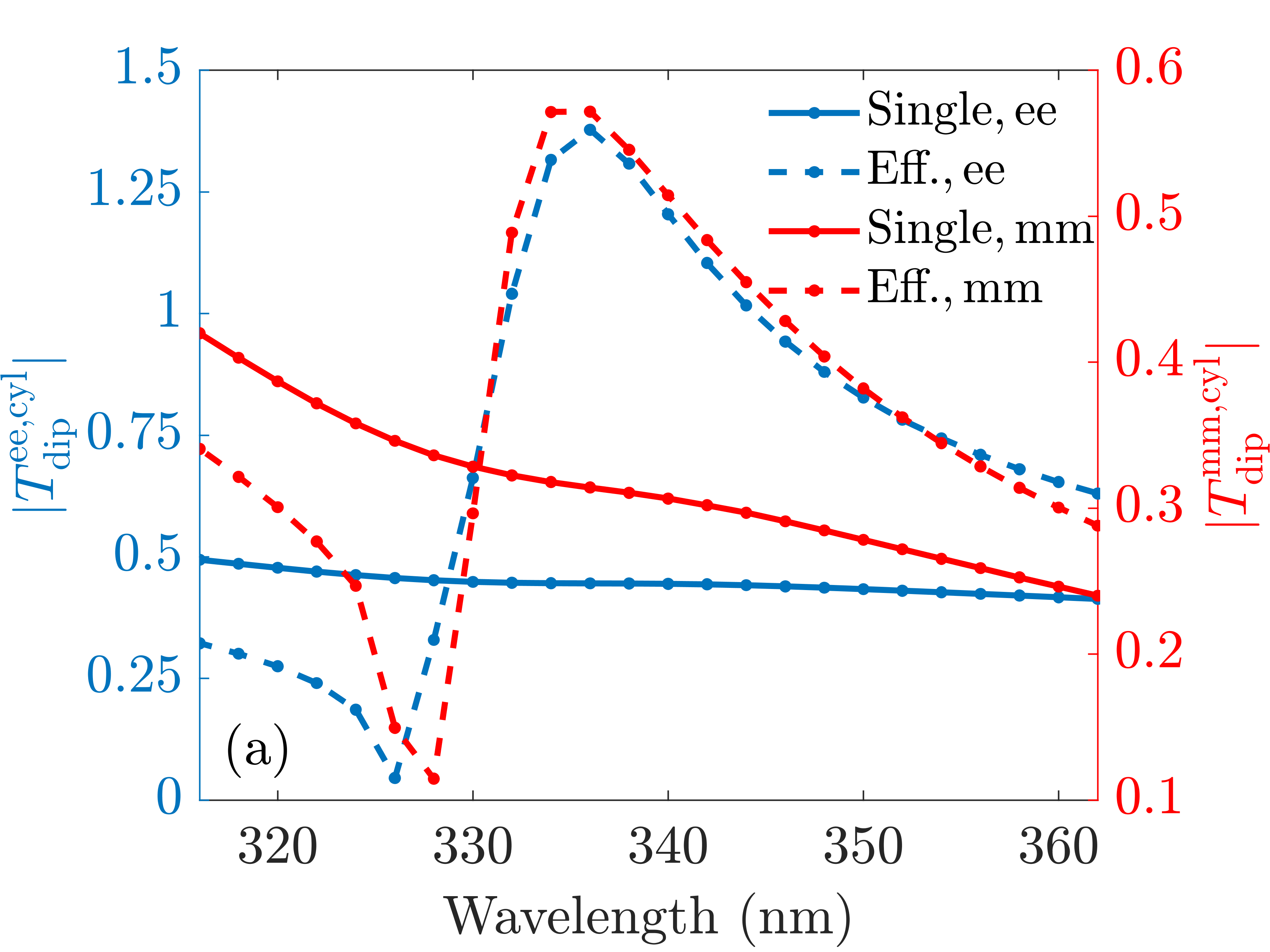}
	}\hspace{1cm}
	\subfloat{
	\includegraphics[width=0.45\textwidth]{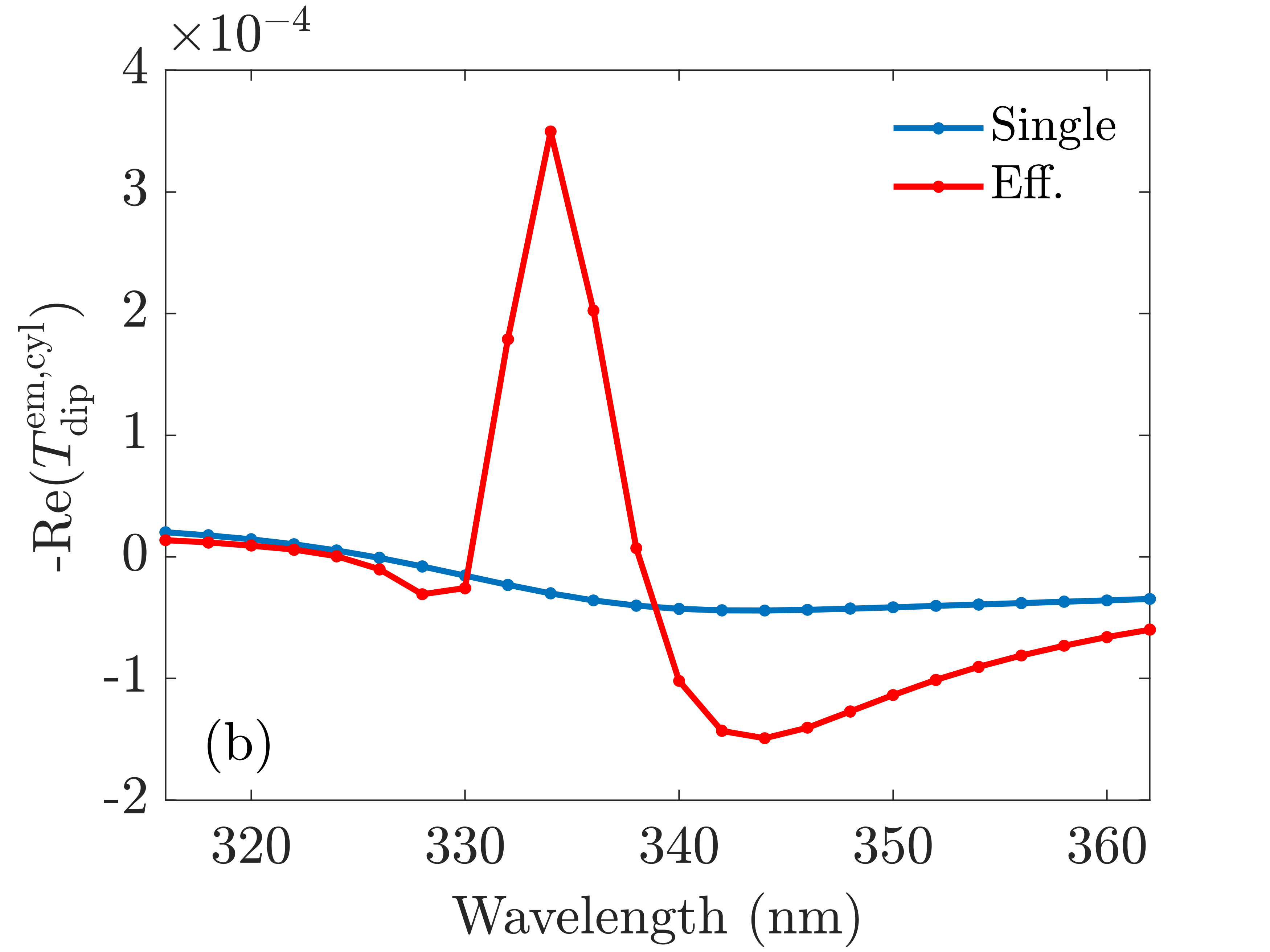}
	}
	\caption{\textbf{(a)} Electric and magnetic dipolar entries of the T-matrix of a single cylinder and of the cylinder inside the two-dimensional lattice, here called the effective T-matrix. \textbf{(b)} Electric-magnetic dipolar entries for the same systems. Both spectra show resonances in all entries of the effective T-matrix close to the design wavelength of 334\,nm.}
    \label{fig:TEEMMEMCyl}
	\end{figure*}

In Figure~\ref{fig:TEEMMEMCyl}\textbf{(a)}, the electric and magnetic dipolar components of the T-matrix of the isolated cylinder and the effective T-matrix of the cylinder inside the lattice are shown. We restrict the consideration to the diagonal components of the T-matrices in the Cartesian basis, which are in the polarization plane of the incident light. In Figure~\ref{fig:TEEMMEMCyl}\textbf{(b)}, the corresponding entries of the electric-magnetic coupling term of the isolated and effective T-matrices are shown. We observe that both the effective electric and magnetic terms show narrow resonances close to 334\,nm. That response is in stark contrast to the response of the single cylinder, for which the spectra is nearly non-dispersive in this frequency range. The electric-magnetic coupling term also shows a considerably more narrow resonance close to 334\,nm for the cylinder in the lattice compared to the single cylinder, which shows a broad resonance around this wavelength as well. Here, we can trace the impact of the lattice on the overall optical response. The structure and the periodic arrangement drastically modify the optical response, and a resonance is observed where the isolated cylinder does not support a resonance. 

In the specific configuration, the lattice constant of the analyzed lattice is close to the design wavelength. In this case, lattice resonances can collectively couple the multipoles sustained by the cylinders. To explain the emergence of lattice resonances and the characterization of the collective coupling of multipoles inside a lattice, see \cite{doi:10.1063/5.0024274,rahimzadegan2021comprehensive}, for instance. The asymmetric Fano resonance observed in this spectral range in the circular dichroism spectra indicates an interference of different competing pathways determining whether the lattice resonance is left- or right-handed and, therefore, also the sign of the circular dichroism, see \cite{doi:10.1063/5.0118263}. Three aspects influence the resonance around 334\,nm in the spectra: (i) The chiral response of the MOF material, (ii) the chiral, electric, and magnetic dipole moments of the isolated MOF cylinders, and (iii) the lattice coupling. The interplay among all these effects, which sensitively depends on the geometry and the intrinsic material parameters, causes even the reversal of the sign of the ACD and the CD signal at a wavelength of 334\,nm. This is a drastic example of how a structure can modify the intrinsic properties of a given material.

The negative real part of the electric-magnetic coupling term of the T-matrix, $-\mathrm{Re}(\mathbf{T}_{\mathrm{em}})$ is proportional to the imaginary part of the chirality. This can be seen by the formulas for the effective parameters in Equation~(17) of \cite{HomogenEffTMat}. The imaginary part of the chirality is the quantity determining the circular dichroism. It shows different signs for the cylinder in the lattice in comparison to the single cylinder and the MOF material, as we show in the Supporting Information in Figure~S5. This indicates that the collective coupling of the multipoles inside the lattice causes the sign change of the circular dichroism of Figure~\ref{fig:CDRBinolCyl}.

An additional observation in the spectra is that the absolute value of the circular dichroism at 334\,nm is not the largest. For wavelengths around 344\,nm, an enhancement of 2.6 and 3.1 is reached for the ACD and CD spectra, respectively. This can be related to the dispersion of the MOF material and geometry-dependent effects. For the MOF material and the slab on the substrate, the transmission circular dichroism shows a maximum at 334\,nm, and the absorption circular dichroism is larger than around 344\,nm. This predestines the enhancement of the circular dichroism at this wavelength since the peak can be measured easily for a slab in an experiment. For a single cylinder, the chirality represented by the electric-magnetic coupling term of the single T-matrix is, however, smaller for this wavelength than for 344\,nm. As the near-fields and the optical chirality density are also strong close to the design wavelength, Equation~(7) in \cite{doi:10.1021/acsphotonics.8b01454} shows that the circular dichroism can be larger in that range than at the design wavelength. 

This example proves that our multiscale approach can be used to design complex photonic devices incorporating molecular structures, such as a metasurface of MOF cylinders in a square lattice showing considerable lattice effects.

\section{Conclusions and Outlook}
We have applied for the first time a multiscale modeling approach to study the electronic circular dichroism of complex molecular materials and structured photonic devices functionalized by them. Our approach bridges multiple length scales by combining precise quantum chemistry simulations of the molecular properties and optical full-wave simulations of periodic materials based on scattering techniques and finite-element method-based solvers. The correctness of our {\it ab initio} calculated structure and optical material properties were confirmed by comparing predictions to experimental results obtained by the X-Ray diffractogram, UV-Vis absorption, and ECD spectroscopy of a thin film of the prepared MOF material. The T-matrices constructed from precise quantum chemistry simulations of such novel material allow us to use them in the description of complex photonic devices. An important step is also the possibility of determining effective material parameters of such materials by applying a dedicated homogenization procedure. 

We demonstrated the predictive power of our multiscale approach at two very distinct photonic devices. One of them was a Fabry-Pérot cavity made from thin aluminum films filled by the crystalline UiO-BINOL material. The second example was a metasurface made from a periodic arrangement of cylinders of the same MOF material arranged in a square lattice. In the former case, we were able to observe regions of high absorption of the cavity modes. These modes are strongly bent due to the interaction of the light with the MOF material within the cavity. In the latter case, we specifically designed a metasurface of finite-size cylinders build from the MOF to resonantly enhance the absorption circular dichroism and transmission circular dichroism signal around a specific wavelength. Since no other materials were present in the simulations, this means that the structured molecular material enhances its own chiral signal. By a careful selection of design parameters, we managed to enhance the ACD signal by a factor of 2.6 and the CD signal by 3.1. 

In general, our approach to modeling photonic devices made from chiral materials is a viable tool for the knowledge-driven computer-aided design and optimization of optical properties and entire devices built with complex materials. It pushes the limits well beyond the phenomenological approach of most current state-of-the-art approaches by explicitly considering the materials from first principles and not just as an adjustable parameter at the level of Maxwell's equations. Since we demonstrated the uniqueness and versatility of our multiscale approach, as an outlook, we could imagine that the presented approach can be utilized in many relevant fields, such as information processing based on circularly polarized light, chiral molecules detection and identification, medicine, or industrial production. Our multiscale modeling approach opens a route to speed-up design and optimization, as well as to reduce the time and costs related to developing complex chiral photonic devices.

\section{Appendix: T-matrix Formalism}
The optical simulations of the multi-scattering problem of the molecular unit cells arranged in a three-dimensional lattice are based on the T-matrix. The T-matrix introduced by Waterman relates the incident to the scattered electromagnetic fields of an object \cite{Waterman1965}. In this article, these objects are molecular unit cells of a MOF. In the frequency domain, the electric field outside an isolated scattering object is defined as
\begin{align}\label{eq:EExpan}
\begin{split}
    \bm{E}(\bm{r})=\sum_{l=1}^{\infty}\sum_{m=-l}^{l}
    \left(
        a_{lm,\mathrm{N}}\bm{N}^{(1)}_{lm}(k_{\mathrm{h}}\bm{r})
        + a_{lm,\mathrm{M}}\bm{M}^{(1)}_{lm}(k_{\mathrm{h}}\bm{r})
    \right.
        \\
    \left.
        + p_{lm,\mathrm{N}}\bm{N}^{(3)}_{lm}(k_{\mathrm{h}}\bm{r})
        + p_{lm,\mathrm{M}}\bm{M}^{(3)}_{lm}(k_{\mathrm{h}}\bm{r})
    \right),
    \end{split}
\end{align}
where $\bm{N}^{(1)}_{lm}(k_{\mathrm{h}}\bm{r})$ and $\bm{M}^{(1)}_{lm}(k_{\mathrm{h}}\bm{r})$ are incident and $\bm{N}^{(3)}_{lm}(k_{\mathrm{h}}\bm{r})$ and $\bm{M}^{(3)}_{lm}(k_{\mathrm{h}}\bm{r})$ are scattered vector spherical waves corresponding to transverse magnetic (TM) and transverse electric (TE) modes in the case of regular waves \cite{Fernandez-Corbaton:2020,Beutel:21}. $k_{\mathrm{h}}^{}=\omega\sqrt{\varepsilon_{\mathrm{h}}(\omega)\mu_{\mathrm{h}}(\omega)}$ is the wave number of the surrounding host material with $\epsilon_{\mathrm{h}}(\omega)$ the permittivity and $\mu_{\mathrm{h}}(\omega)$ the permeability of the host material at the considered frequency $\omega$. In the following, we skip the frequency $\omega$ as an argument for brevity. The multipolar order $l$ is bounded by a maximum value in simulations. For molecular scatterers, $l=1$ sufficiently describes the response of the object in most cases, as molecules are usually very small compared to the wavelength of the incident waves. The expansion coefficients $a_{lm,\mathrm{N}}$ and $a_{lm,\mathrm{M}}$ of the incident wave and the expansion coefficients $p_{lm,\mathrm{N}}$ and $p_{lm,\mathrm{M}}$ of the scattered wave are arranged in column vectors $\bm{a}$ and $\bm{p}$. The T-matrix relates these vectors, \begin{align}\label{eq:DefTMat}
    \bm{p}
    =\mathbf{T}
    \bm{a}.
\end{align} 
The dipolar T-matrix of a molecular unit can be calculated from its dipolar polarizability tensors via simple matrix multiplications, see Equation~(6) from \cite{Fernandez-Corbaton:2020}:
\begin{align}\label{eq:PolToTMat}
\begin{split}
\begin{pmatrix}
\mathbf{T}_{\mathrm{NN}}&\mathbf{T}_{\mathrm{NM}}\\
\mathbf{T}_{\mathrm{MN}}&\mathbf{T}_{\mathrm{MM}}
\end{pmatrix}
=&\frac{\mathrm{i}c_{\mathrm{h}}Z_{\mathrm{h}}k_{\mathrm{h}}^3}{6\pi}\\
&\begin{pmatrix}
\mathbf{C}\left(\bm{\alpha}_{\mathrm{ee}}\right)\mathbf{C}^{-1}&\mathbf{C}\left(-\mathrm{i}\bm{\alpha}_{\mathrm{em}}/Z_{\mathrm{h}}\right)\mathbf{C}^{-1}\\
\mathbf{C}\left(\mathrm{i}\bm{\alpha}_{\mathrm{me}}/c_{\mathrm{h}}\right)\mathbf{C}^{-1}&\mathbf{C}\left(\bm{\alpha}_{\mathrm{mm}}/(c_{\mathrm{h}}Z_{\mathrm{h}})\right)\mathbf{C}^{-1}
\end{pmatrix}.
\end{split}
\end{align}
In Equation~(\ref{eq:PolToTMat}), $c_{\mathrm{h}}^{}=1/\sqrt{\varepsilon_{\mathrm{h}}^{}\mu_{\mathrm{h}}^{}}$ is the speed of light in the surrounding medium assumed to be achiral and $Z_{\mathrm{h}}^{}=\sqrt{\mu_{\mathrm{h}}^{}/\varepsilon_{\mathrm{h}}^{}}$ is the wave impedance of the surrounding medium \cite{Fernandez-Corbaton:2020}. $\mathbf{C}$ is a unitary matrix that converts the polarizability tensors from the Cartesian to the spherical basis. The complex $3\times 3$ dipolar polarizability tensors $\bm{\alpha}_{\nu \nu'}$ can be computed using \textit{ab initio} quantum mechanical methods as TD-DFT \cite{Fernandez-Corbaton:2020,SURMOFCavity}. Note that the T-matrix defined by Equation~(\ref{eq:PolToTMat}) is in the basis of well-defined parity. Considering chiral scatterers like chiral molecules, choosing the basis of well-defined helicity is more convenient. One can transfer the T-matrix from the basis of well-defined parity into the basis of well-defined helicity via simple matrix multiplications as defined in Equation~(S47) in \cite{Fernandez-Corbaton:2020}.

In a MOF, the molecular units are arranged in a lattice representing a multi-scattering problem between the unit cells. The solution to this multi-scattering problem in two dimensions reads \cite{Beutel:21}
\begin{align}\label{eq:MultiScat}
    \bm{p}_{0,\mathrm{tot}}=\left(\mathds{1}-\mathbf{T}\sum_{\bm{R}\neq 0}\mathbf{C}^{(3)}(-\bm{R})\mathrm{e}^{\mathrm{i}\bm{k}_{\parallel}\bm{R}}\right)^{-1}\mathbf{T}\bm{a}_0,
\end{align}
where $\bm{a}_0$ are the expansion coefficients of the primary incident field and $\bm{p}_{0,\mathrm{tot}}$ are the expansion coefficients of the total field scattered by an object at the origin of the lattice. $\mathbf{C}^{(3)}(-\bm{R})$ is a matrix of translation coefficients for vector spherical waves, $\bm{R}$ is a lattice point, and $\bm{k}_{\parallel}$ is the tangential component of the primary illuminating plane wave. The wave vector-dependent effective T-matrix of a two-dimensional lattice, defined as
\begin{align}\label{eq:Teff2D}
    \mathbf{T}_{\mathrm{eff}}=\left(\mathds{1}-\mathbf{T}\sum_{\bm{R}\neq 0}\mathbf{C}^{(3)}(-\bm{R})\mathrm{e}^{\mathrm{i}\bm{k}_{\parallel}\bm{R}}\right)^{-1}\mathbf{T},
\end{align}
describes the scattering from an object at the origin of the lattice incorporating the multi-scattering problem. Note that this effective T-matrix is defined for a two-dimensional lattice and depends on the wave vector of the incident light wave. Therefore, it differs from the effective T-matrix defined in Equation~(3) of Ref.~\cite{HomogenEffTMat}, which is defined for a three-dimensional lattice and does not depend on the propagation direction of the incident light. For the calculation of the effective material parameters of the MOF, the effective T-matrix from Equation~(3) of Ref.~\cite{HomogenEffTMat} is used. 

In the in-house developed code mpGMM, Equation~(\ref{eq:MultiScat}) is solved, and the response of a multi-layered system is calculated by considering the interaction of stacked identical two-dimensional lattices and/or isotropic and homogeneous slabs \cite{Beutel:21}. This code is used in this article to compute the absorption and circular dichroism of the macroscopic MOF film.

% Acknowledgements
\medskip
\textbf{Acknowledgements} \par %delete if not applicable))
M.K., D.B., S.O., S.B., C.W., and C.R. acknowledge support by the Deutsche Forschungsgemeinschaft (DFG, German Research Foundation) under Germany’s Excellence Strategy via the Excellence Cluster 3D Matter Made to Order (EXC-2082/1-390761711) and from the Carl Zeiss Foundation via the CZF-Focus@HEiKA Program. M.K., C.H., and C.R. acknowledge funding by the Volkswagen Foundation. I.F.C., C.W., and C.R. acknowledge support by the Helmholtz Association via the
Helmholtz program “Materials Systems Engineering” (MSE). B.Z. and C.R. acknowledge support by the KIT through the “Virtual Materials Design” (VIRTMAT) project. S.O. achnowledge the ‘Fonds der Chemischen Industrie’ for the scholarship. LH. and S.B. acknowledge the funding by the Deutsche Forschungsgemeinschaft (DFG, via SPP 1928 COORNETs and via HE 7036/5). C.L. thanks the China Scholarship Council (CSC) for financial support. We thank Dr. Markus Nyman for fruitful discussions and insightful comments on our work. M.K. and C.R. acknowledge support by the state of Baden-Württemberg through bwHPC and the German Research Foundation (DFG) through grant no. INST 40/575-1 FUGG (JUSTUS 2 cluster) and the HoreKa supercomputer funded by the Ministry of
Science, Research and the Arts Baden-Württemberg and by the Federal Ministry of Education and Research. We are grateful to the company JCMwave for their free provision of the FEM Maxwell solver JCMsuite. 

%Open access funding enabled and organized by Projekt DEAL.

\medskip
\textbf{Conflict of Interest}
The authors declare no conflict of interest.

\medskip
\textbf{Data Availability Statement}
The data that support the findings of this study are available from the
corresponding authors upon reasonable request. Additionally, DFT simulations produced within this work are available at NOMAD materials database under following DOIs: 10.17172/NOMAD/2022.11.30-1 and \\10.17172/NOMAD/2022.11.30-2. 
The Supporting Information covers detailed material on the conducted experiments and their results. Data that refers to the herein described experiments were submitted to the repository Chemotion (\url{https://www.chemotion-repository.net/}). All DOIs minted for the data are linked to the specific experiments in the Supporting Information and a summary of all new data obtained in this study can be gained with the collection DOI \url{https://dx.doi.org/10.14272/collection/SiO\_2023-01-18}.\cite{ChemotionSimon}

\bibliography{bibliographyArxiv}

\end{document}